\documentclass[useAMS,usenatbib,usegraphicx]{mn2e}

\usepackage{threeparttable}
\usepackage{multirow}

\newcommand{\SII}{[S~{\sc ii}]\ }
\newcommand{\NII}{[N~{\sc ii}]\ }
\newcommand{\HII}{H~{\sc ii}\ }
\newcommand{\HI}{H~{\sc i}\ }
\newcommand{\Ha}{H$\alpha$\ }
\newcommand{\kms}{\,\mbox{km}\,\mbox{s}^{-1}}
\newcommand{\HST}{\textit{HST}\ }
\newcommand{\SIIHa}{I([S~{\sc ii}])/I(H$\alpha$)}
\newcommand{\NIIHa}{I([N~{\sc ii}])/I(H$\alpha$)}
\newcommand{\OIIIHb}{I([O~{\sc iii}]5007)/I(H$\beta$)}

\begin{document}

\title[SGS in IC~2574: gas kinematics.]{The supergiant shell with triggered star formation in Irr galaxy IC~2574: neutral and ionized gas kinematics.}

\author[Egorov et al.]{
   O.V.~Egorov$^{1}$\thanks{E-mail: egorov@sai.msu.ru},
   T.A.~Lozinskaya$^{1}$,
   A.V.~Moiseev$^{1,2}$,
   and G.V.~Smirnov-Pinchukov$^{1}$ \\
$^{1}$ Lomonosov Moscow State University, Sternberg Astronomical Institute,
        Universitetsky pr. 13, Moscow 119991, Russia
       \\
  $^{2}$ Special Astrophysical Observatory, Russian Academy of Sciences\thanks{The system of the Russian Academy of Sciences institutes was liquidated in Sep 2013}, Nizhnii Arkhyz 369167, Russia
   }

\date{Accepted 2014 July 6. Received 2014 June 18; in original
form 2014 April 5}

\pagerange{\pageref{firstpage}--\pageref{lastpage}} \pubyear{2014}

\maketitle

\label{firstpage}

\begin{abstract}

We analyse the ionized gas kinematics in the star formation regions of the supergiant shell (SGS) of the IC~2574 galaxy using observations with the Fabry--Perot interferometer at the 6-m telescope of SAO RAS; the data of the THINGS survey are used to analyze the neutral gas kinematics in the area. We perform the `derotation' of the \Ha and \HI data cubes and show its efficiency in kinematics analysis. We confirm the SGS expansion velocity 25 $\kms$ obtained by \citet{walter99} and conclude that the SGS is located at the far side of the galactic disc plane. We determine the expansion velocities, kinematic ages, and the required mechanical energy input rates for four star formation complexes in the walls of the SGS; for the remaining ones we give the limiting values of the above parameters. A comparison with the age and energy input of the complexes' stellar population shows that sufficient energy is fed to all \HII regions  except one. We discuss in detail the possible nature of this region and that of another one, which was believed to be an SNR according to radio observations. We measured the expansion velocity  of the latter and confirm its identification as an old SNR.
Our observations allowed us to identify a faint diffuse \Ha emission inside the SGS which was never observed before.

\end{abstract}


\section{Introduction.}

Extensive observations of $1-2$~kpc large supergiant shells (SGS) and holes in the \HI distribution in
galaxies have been performed for a long time  \citep[see e.g.][]{Warren11}.
 Giant holes in  some dwarf irregular (dIrr) galaxies represent
the dominant feature of the ISM \citep*[see e.g.][and references therein]{Young97, ott01, Simpson05, cannon11a, Warren11} and their origin has been the subject of debate for more than two decades.
   In the standard approach based on the \citet{weaver77} model, \HI shells
result from the cumulative action of multiple stellar winds and supernovae
explosions  \citep[see e.g.][]{mccray87, tenor88, ott01}.
However, it was recognized long ago \citep[see, e.g.,][]{tenor88, rhode99, kim99,  Simpson05, silich06} that
this scenario cannot explain the origin of giant supershells in which the mechanical energy input from the detected stellar clusters appears to be  inconsistent with that required by the standard  model.
 Several other mechanisms have been proposed for the
formation of the observed \HI distribution and kinematics in galaxies: collisions of high velocity clouds with galactic
discs \citep{tenor81}, radiation pressure from field
stars \citep{elmegreen82}, non-linear evolution of
self-gravitating turbulent galactic discs \citep*{wada00, dib05}, ram pressure of the intergalactic medium
\citep{bureau02}, fractal ISM \citep{elmegreen97}, \HI dissolution by UV radiation \citep{vorobyov04}, and even exotic mechanisms (see references in \citealt{silich06} and \citealt{Warren11}).

\begin{figure}
\includegraphics[width=\linewidth]{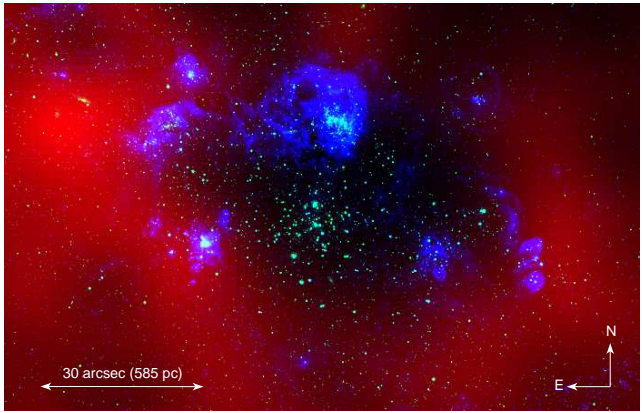}
\caption{False-colour image of the SGS in IC~2574. Red -- \HI column density map, blue -- \Ha+ \NII intensity from \textit{HST}/ACS F658N image, green -- continuum from \textit{HST}/ACS F814W image. The location of the SGS on the IC~2574 \HI map is shown in Fig.~\ref{fig:velmodel}.}\label{fig:SGS-color}
\end{figure}

Recent studies have found that multiple star formation events over the lifetime of the giant hole do provide enough
energy to drive the hole formation. The situation has become clear as a result of \textit{Hubble Space
Telescope (HST)} observations.
 \citet {weisz09a} showed that the holes in the Holmberg~II galaxy contain multiple stellar generations and could indeed be formed from the energy input of these stars. Today it is clear that high star-formation rates can be maintained for a much longer time period than previously believed: dwarf galaxies are characterized by the ability to sustain a high
star-formation efficiency over several hundred Myr, and even up to
1000~Myr in some cases \citep{mcquinn09,mcquinn10a,mcquinn10b}, with local short starbursts occurring during this
period. Such long periods of intense star formation provide enough energy from stellar winds and supernova
explosions to drive the formation of giant cavities and supergiant shells
\citep[see e.g.][]{weisz09a,weisz09b,cannon11a,cannon11b}. Note that a number of SGSs exhibit conspicuous
signs of expansion-triggered star formation at their periphery.
Of special interest is a detailed analysis of the interaction of stars and gas in the regions
of new star formation sites in the walls of giant shell-like H~{\sc i} structures, which can elucidate
the process of their evolution.

Dwarf irregular galaxies provide the best environment for studying the  creation mechanisms of giant \HI
structures with star-formation episodes on their rims.
Because of the slow solid-body rotation and the lack of strong spiral density waves which can destroy the giant shells,
they grow to a larger size and live longer compared to such structures in spiral galaxies. The overall
gravitational potential  of dwarfs is much smaller than that of spiral galaxies, their \HI disc scale
height is larger and the gas volume density is lower than in spirals. Therefore the same amount of
mechanical energy fed to the ISM of dwarf Irr galaxies creates very large long-lived holes with star
formation in the walls triggered by their expansion.

The aim of this work is to clarify the effect on supergiant HI shells of new bursts of star formation in their walls. For this purpose  we analyse the recent bursts of star formation in the rims of  the  well known
'supergiant shell'  in the dIrr galaxy IC~2574
 based on the detailed kinematics of both neutral and ionized gas.

The IC~2574 galaxy (alternative names: UGC~5666, DDO~81, VII~Zw~330), which is a  member of the M81 group,
is  located at a distance $D=4.02$~Mpc (1~arcsec = 19.5~pc) according to \citet{karach13}; the systemic velocity of
IC~2574 is $V_{hel}\simeq 53 \kms$. It hosts a large number of giant \HI supershells and holes:  48 holes have
been identified by \citet{walter99}, 27 holes --  by \citet{bagetakos11}; see also \citet{rich08}. The
difference in the total number of holes is due to the different criteria used to classify a structure as a
hole or a supershell.

The  SGS surrounding a $1000\times500$ pc hole in the most prominent region of current and recent
star-formation activity  near the N-E outskirts of the IC~2574 galaxy represents an impressive example of a
giant \HI shell with triggered SF along the rims (see Fig.~\ref{fig:SGS-color} and Fig.~\ref{fig:velmodel}). It is the best candidate for a detailed study of the interplay between the
recent bursts of star formation and the gas in the rims of giant supershells. This  elliptical SGS denoted
as \#~35 in the list of \citet{walter99} has been extensively studied at radio \citep{walter98, walter99},
UV \citep{stewart00}, optical \citep{tomita98, pasquali08, weisz09b}, mid-infrared \citep{cannon05, dalcanton12}, and
X-ray  wavelengths \citep[see][and references therein]{walter98, yukita12}.

An analysis of the resolved stars based on \textit{HST} data
\citep{weisz09b} shows that the last most significant
episodes of star formation in the SGS began $\simeq$ 100~Myr ago and
the recent bursts of star formation along its walls  are as young as
 $\le 10$ Myr. The ages of the younger star-formation
events are consistent with those derived from broadband
photometry (e.g., \citealt{stewart00, cannon05, pasquali08}) and are younger
than the estimated kinematic age of the \HI shell, $14 \pm 3$~Myr \citep{walter98, walter99}.

In this paper we present a new study of the morphology  and kinematics of ionized and neutral gas in the  SGS. Our analysis is based on observations made with the 6-m telescope of the Special Astrophysical
Observatory (SAO RAS). We also use the Very Large Array (VLA)  21 cm  observations of \HI gas in the galaxy from the THINGS survey \citep{things}.

The paper is organized as follows:
in Section~\ref{sec:obs} we describe the observations and data reduction.
Section~\ref{sec:HI} presents the results of our re-estimation of the SGS  expansion velocity from the 21~cm data.
Section~\ref{sec:HII} presents an analysis of ionized  gas kinematics in the star-formation
regions in the rims of the SGS.
In Section~\ref{sec:HII-results} we discuss the obtained results. Section~\ref{sec:conclusions} summarizes our main conclusions.

\section{Observations and data reduction}\label{sec:obs}

 \subsection{Optical FPI-observations}

\begin{table*}
\caption{Properties of observational data}
\label{tab:obs_data}
\begin{tabular}{llllllll}
\hline
Data set            & Date of obs & $\mathrm{T_{exp}}$  & FOV &  $''/px$  & $\theta$, $''$ &  sp. range  & $\delta\lambda$ or $\delta v$ \\
\hline
\multirow{2}*{scanning FPI} & 17/18 Oct 2012 & $40\times120$ sec &   \multirow{2}*{$6.1'\times6.1'$} & \multirow{2}*{0.71}  &  \multirow{2}*{2.1} & \multirow{2}*{8.8\AA\, around \Ha} & \multirow{2}*{0.48\AA \,($21\kms$) } \\
                                            & 18/19 Oct 2012 & $40\times120$ sec  &     &     &   &  &\\
THINGS \HI           & 03 Sep 1992      &    14 h &   $25.6'\times25.6'$   &  1.5  & $12.8\times11.9$ & $261\kms$ around 21 cm  & $2.6 \kms$\\
FN655 image        & 07/08 Mar 2013 & 1800 sec                   & $6.1'\times6.1'$                       & 0.36                      &  2.0 &    \Ha$+$\NII & \\
FN674 image        & 07/08 Mar 2013 & 3000 sec                   & $6.1'\times6.1'$                       & 0.36                      &  2.0 &     \SII & \\
long-slit               & 19/20 Nov 2012  & 3600 sec                 & $6.1'\times1.0''$                           & 0.35                     & 1.7  & 5800--7400\AA  & 5.5\AA \\
F658N  image       &  6 Jan 2008                       &   1200 sec                      &   $3.4'\times3.4'$                                             &   0.05                           & 0.09    &    \Ha$+$\NII & \\
F814W  image       &  6 Jan 2008                &    6400 sec                      &   $3.4'\times3.4'$                                             &    0.05                          & 0.10    &    continuum & \\
  \hline
\end{tabular}
\end{table*}

The observations were made at the prime focus of the 6-m
telescope of SAO RAS using a scanning Fabry--Perot interferometer (FPI) mounted  inside the
SCORPIO-2  multi-mode focal reducer \citep{scorpio2}. The operating spectral range  around the \Ha emission line was
cut by a narrow bandpass filter with a $\mathrm{FWHM}\approx14$~\AA\ bandwidth. The FPI751 interferometer
provides a free spectral range between the neighbouring interference orders $\Delta\lambda=8.8$~\AA\, with
a spectral resolution (FWHM of the instrumental profile) of about $0.48$~\AA. During the scanning process, we have consecutively obtained 40 interferograms  at different distances
between the FPI plates. The log of these observations and the parameters of other data sets are given in Table~\ref{tab:obs_data}, where $\mathrm{T_{exp}}$ is the exposure time, FOV -- the field of viev,  $\theta$ -- the final angular resolution, and $\delta\lambda$ or $\delta v$ is the final spectral or velocity resolution.

The data reduction was performed using a software package running in the \textsc{idl} environment. After the initial reduction, sky line subtraction, and photometric and seeing corrections made using the reference stars
and wavelength calibration, the observational data were combined into data cubes, where each pixel in
the field of view contains a 40-channel spectrum. (For a detailed description of the data reduction algorithms
and software see \citealt{Moiseev02ifp} and \citealt{MoiseevEgorov2008}.)
We observed the galaxy at two position angles in order to remove the parasitic ghost reflection. These data were reduced separately to get the wavelength cubes of
the object. Both cubes were then co-added.

In order to get rid of the stellar continuum emission in the constructed data cube and reveal only the gas emission details, we perform a second-order background fit in each pixel at the edges of the line profiles and subtract it.  The analysis of the \Ha line profiles was carried out using the multi-component Voigt fitting  \citep{MoiseevEgorov2008}.

The channel maps of data cube obtained shown in Fig.~\ref{fig:Ha-channel} in Appendix.

\subsection{Narrow-band imaging}

The deep optical images in the \Ha and \SII emission lines were taken at the prime focus of the 6-m telescope of SAO RAS with SCORPIO-2 multi-mode focal reducer using filters FN655 and FN674 with  the central  wavelengths   6559 and 6733 \AA\,  and FWHM = 97 and 60 \AA\, respectively. While the FWHM of the FN655 filter is broader than the distance  between the \Ha and [N~{\sc ii}]\, emission lines, the image in  this filter is contaminated  by [N~{\sc ii}] 6548, 6584~\AA\, emission. The image obtained with FN674 represents the emission
of the \SII 6717, 6731~\AA\, lines.

We used the broader-band  FN608 and i-SDSS filters centered on the continuum near the
two emission lines  to subtract the stellar contamination from the images obtained on the same night.
Note that the subtraction was not ideal and residuals due to the stellar contribution can be seen in our final images in several regions, especially in those taken in the \SII
lines.  In order to calibrate the emission-line images to energy fluxes we observed the
standard star AGK+81d266 immediately after observing the galaxy.

\begin{figure}
\includegraphics[width=0.48\linewidth, angle=90]{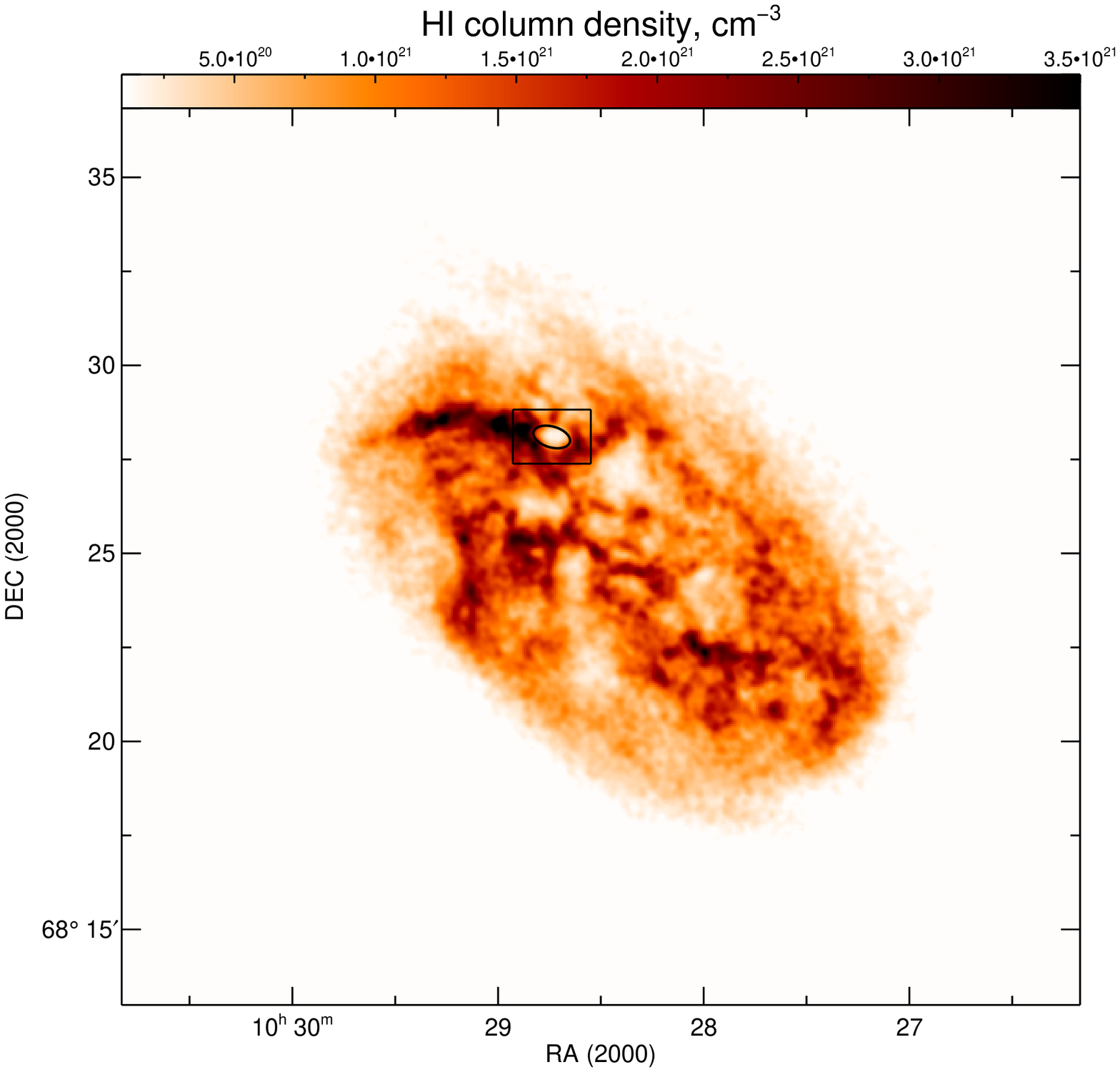}
\includegraphics[width=0.48\linewidth, angle=90]{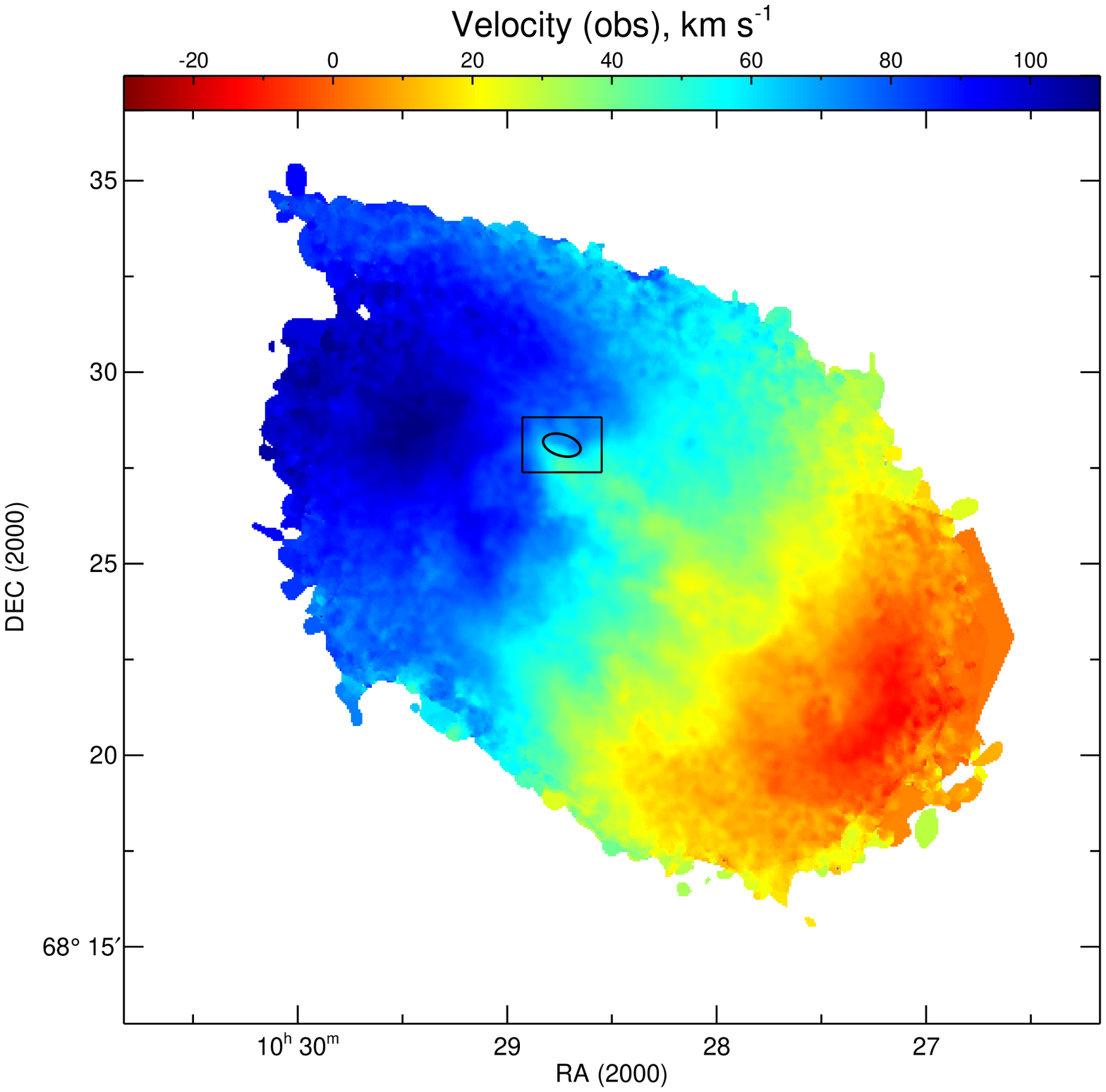}

\includegraphics[width=0.48\linewidth, angle=90]{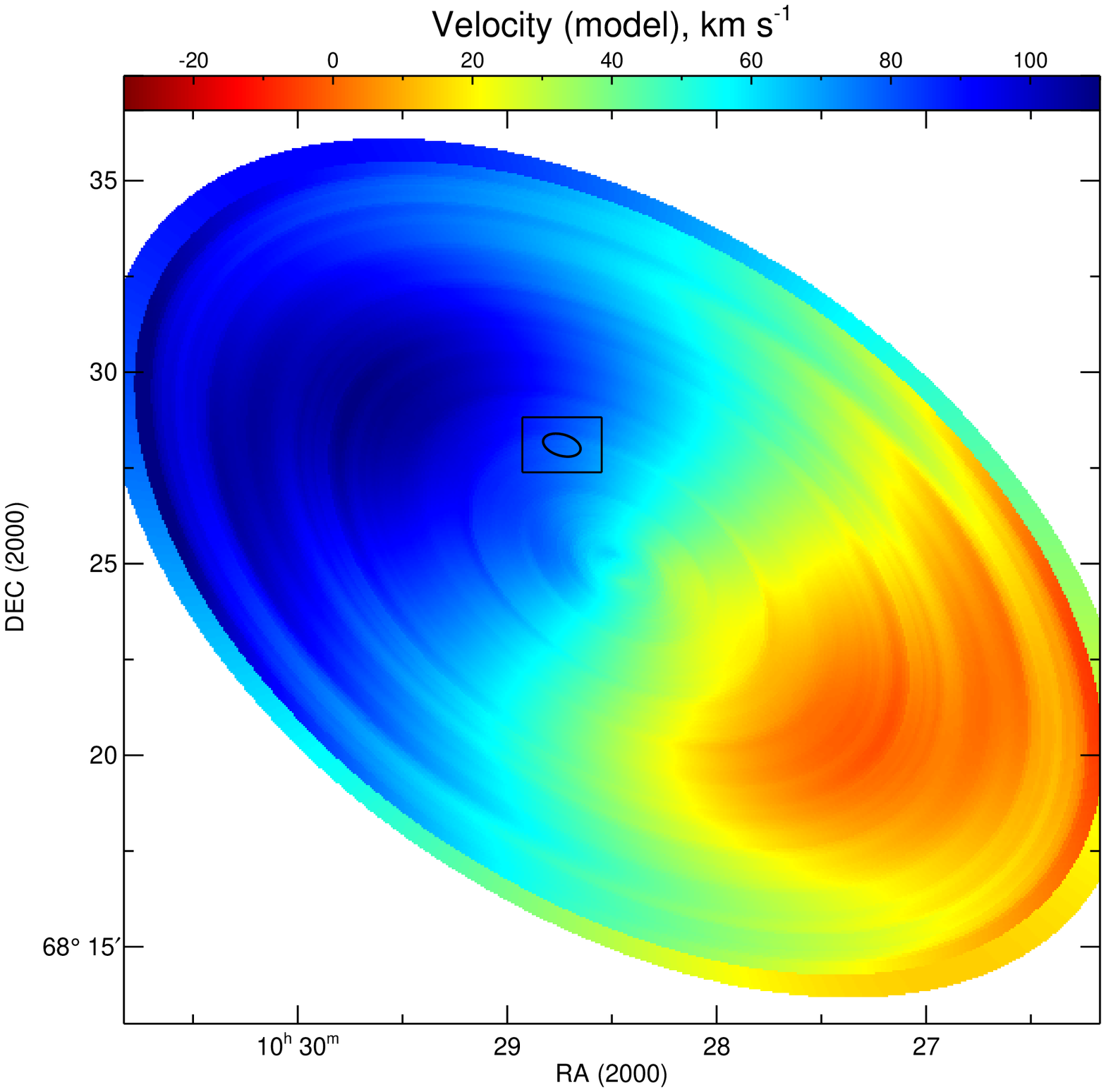}
\includegraphics[width=0.48\linewidth, angle=90]{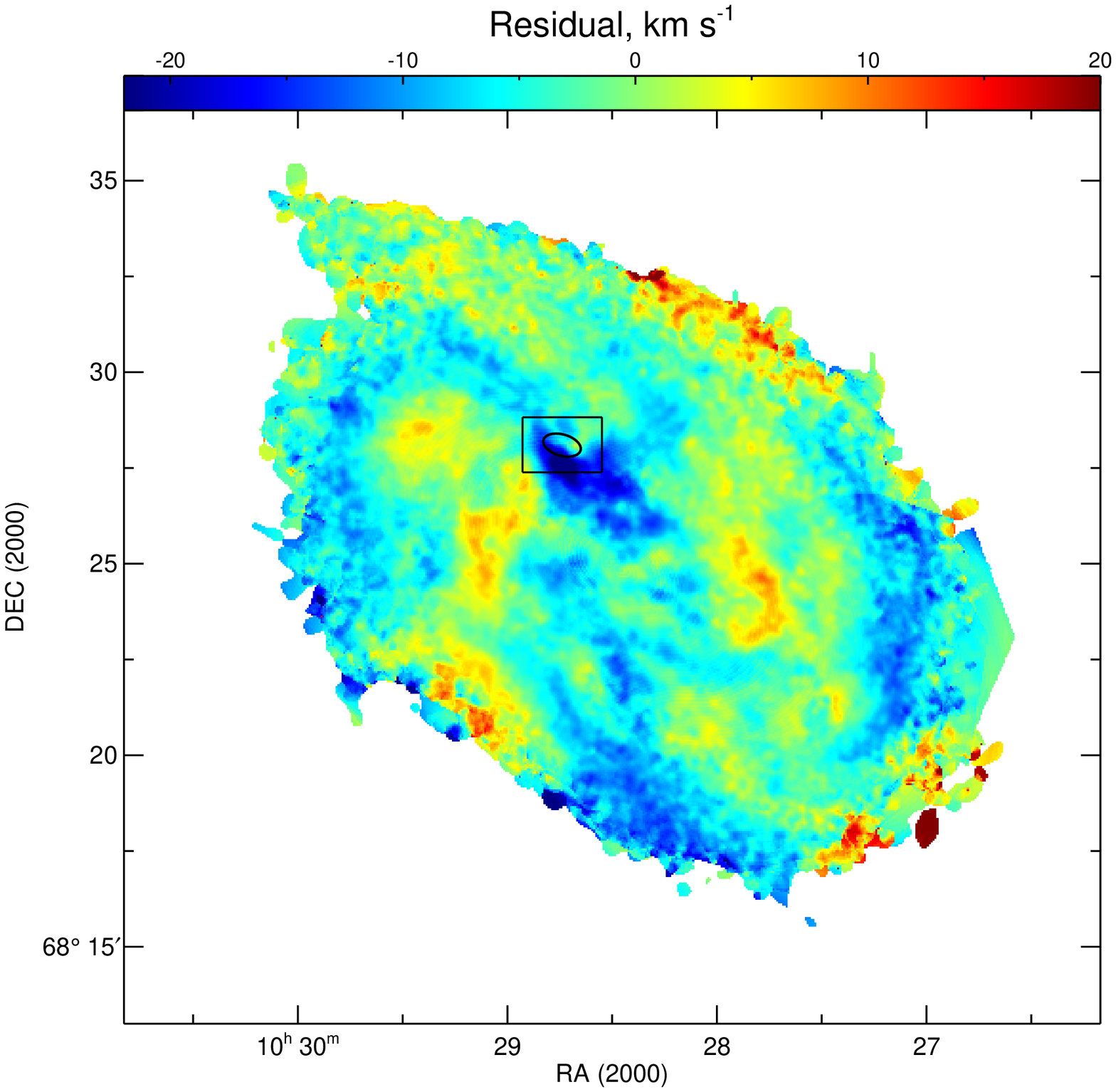}

\caption{IC~2574 in \HI 21~cm: \HI column density map (top left), observed line-of-sight velocity map (top right), tilted-ring model of the velocity field (bottom left), and residuals after its subtraction from the observed \HI velocity field (bottom right). The black ellipse denotes the location of the hole inside the SGS; the black rectangle represents the SGS area shown in Fig.~\ref{fig:SGS-color}.}
\label{fig:velmodel}
\end{figure}

 \subsection{Other observational data}
\label{sec_obs3}

In order to study the \HI gas kinematics in the SGS,  we analysed archival \HI 21 cm line THINGS survey VLA data \citep{things}. In this work we used the
natural-weighted  data cube (see the parameters in Table~\ref{tab:obs_data}).

We performed long-slit spectroscopy of the \HII regions in IC~2574 with the 6-m telescope of SAO RAS and the
SCORPIO focal reducer \citep{scorpio}. An analysis of these data
will be presented in our forthcoming work (Egorov et al., in preparation);
here we use the data obtained with one slit position  to analyse the \SIIHa\ ratio distribution in \HII Region \# 7 (see Sec.~\ref{sec:regEnergy}).

\begin{figure}
\includegraphics[width=\linewidth]{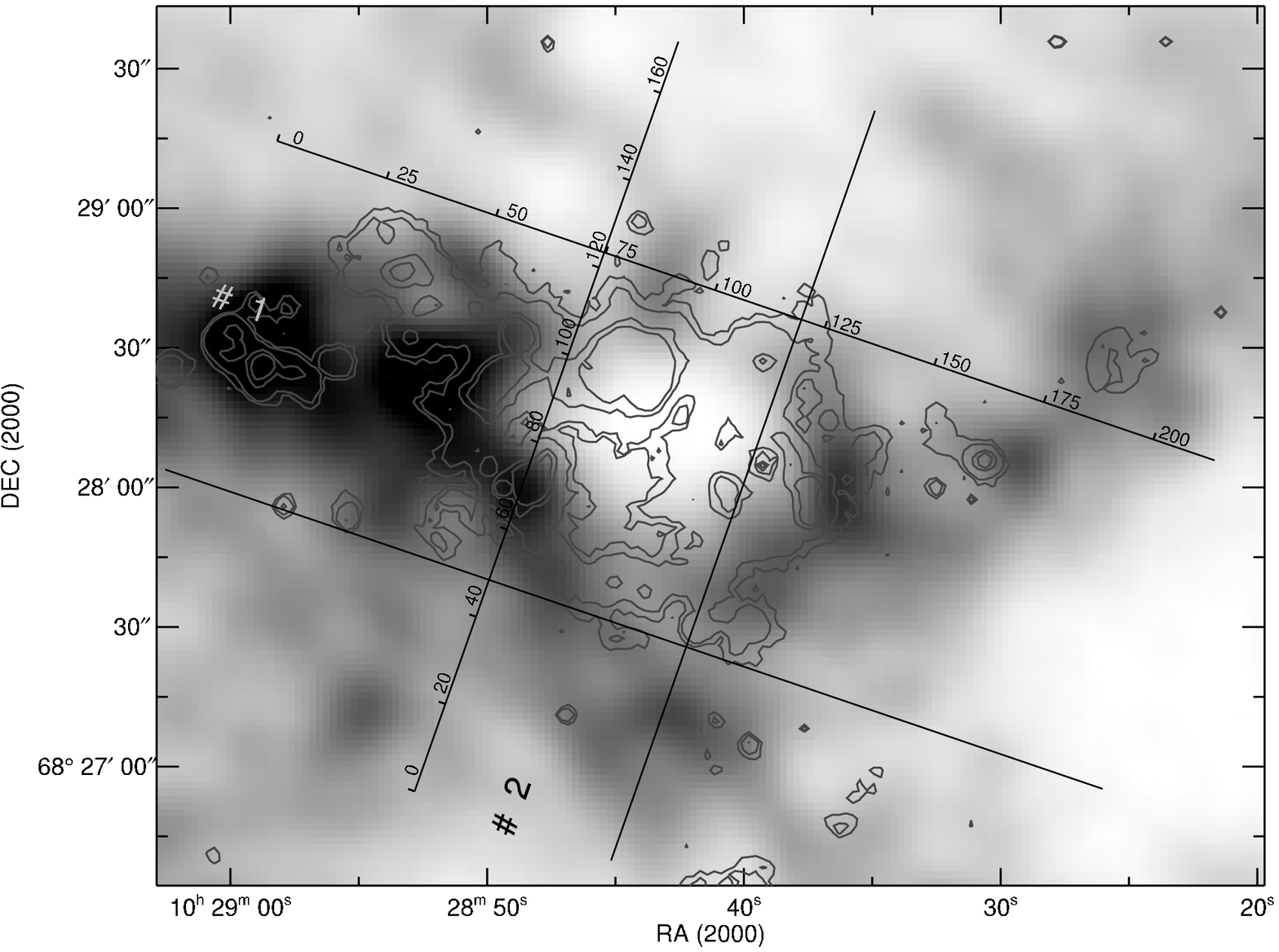}

\includegraphics[width=\linewidth]{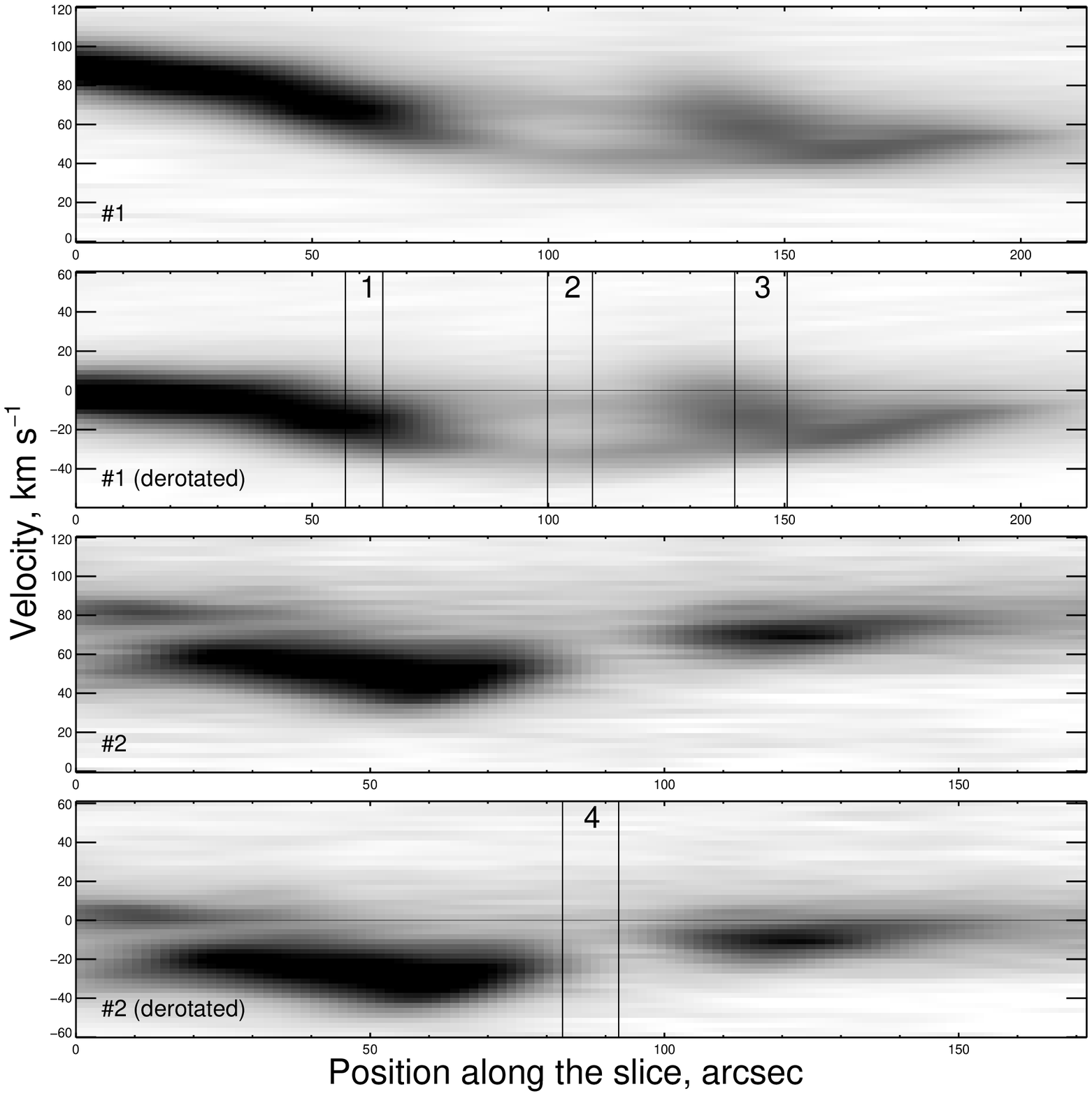}

\caption{Location of the wide PV diagrams on the \HI map of the SGS. The
contours correspond to the \Ha emission intensity levels of
$[0.5, 1.0, 5, 10]\times10^{-16} \mathrm{erg\ s^{-1} cm^{-2}}$  (top panel). Below this image the corresponding \HI 21 cm line PV diagrams are shown both uncorrected and corrected (derotated) for the rotation of the galaxy. The vertical lines on the PV diagrams indicate the location of the profiles shown in Fig.~\ref{fig:thick-prof}.
}
\label{fig:thick-pv}
\end{figure}

To study  the ionized gas morphology in the SGS region we used the archival  \HST Advanced Camera for Survey (ACS) images in the
F658N filter (which contains the H$\alpha$ + [N~\textsc{ii}] emission lines) and the F814W filter presented in \citet{weisz08, weisz09b}.

\begin{figure}
\includegraphics[width=1\linewidth]{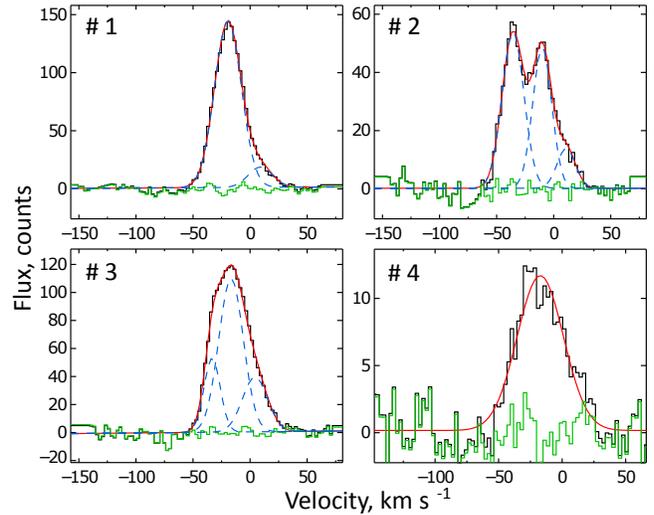}
\caption{\HI 21 cm line profiles and their Gaussian decomposition. The locations of these profiles are shown on the PV diagrams in Fig.~\ref{fig:thick-pv}. The black solid line denotes the observed data, the red solid line -- the fitted profile, blue dashed line -- single Gaussian components, green line -- the residuals.}\label{fig:thick-prof}
\end{figure}

\subsection{Construction of a circular rotation model}\label{sec:rot-model}

The structure and kinematics of gas in the region around the SGS are very complex and it would
therefore be useful to exclude the regular component associated with the rotation of the galaxy to
understand its local kinematics.
We construct a rotation model of IC~2574 from the \HI velocity field (moment 1) using the tilted-ring approximation adapted to dwarf galaxy kinematics  \citep{Moiseev2014}.
The coordinates of the dynamical centre, the inclination and the initial value of the major axis position
angle were adapted from \citet{Oh08}, who performed a careful study of the \HI kinematics of this galaxy. The
model takes into account the radial variation of the position angle of the kinematic major axis in order to
describe the large-scale
non-circular streaming motions associated with the bar and two-armed spiral structure. The final model is based only on the data for the regions whose velocity does not deviate significantly from the initial model based on the entire field. Figure~\ref{fig:velmodel} shows the model  velocity map as well as the observed velocity field, residual velocities and \HI column density map.

We correlated the observed data cubes as follows. The velocity of each spectral channel in each pixel was modified by the value of the radial velocity at a given location in the galaxy taken from the rotation model. A new velocity grid (common for all pixels) was then constructed with the same velocity sampling. The spectrum from each spatial element was then interpolated (`derotated') to the new velocity grid and a new data cube was constructed. This procedure was carried out for the \HI 21cm and \Ha line data cubes presented below.
To demonstrate the result of this correction we show two position -- velocity (PV) diagrams along and across the SGS in Fig.~\ref{fig:thick-pv}, where the corrected PV diagrams denoted as `derotated'. The effect of `derotation' is most clearly seen on PV diagram \# 1 since it crosses the SGS in the direction of significant line-of-sight velocity gradient. Considering the relatively large size of the investigated area and hence a possible large contribution of galactic rotation to the PV diagrams, the `derotation' allowed us to eliminate the effect of the disc rotation and to emphasize the local kinematics.

\section{Global gas kinematics of the SGS }\label{sec:HI}

Two reasons prompted us to re-estimate the  expansion velocity $v_\mathrm{exp}$ of the \HI SGS.
First:  \citet{walter99} measured $v_\mathrm{exp} = 25 \kms$  (see also \citealt{walter98}). However,  \citet{weisz09b} point out that `\citet{walter99} measure an \HI expansion velocity of $25~\kms$  from a break in the PV diagram, which is only an indirect measure
of expansion, and in fact is consistent with stalled expansion as well'.
Second: the neighbourhood of the SGS is a dynamically very complex region of the galaxy in the 21 cm line,
making it difficult to distinguish the effect of SGS expansion.

\begin{figure}
\includegraphics[width=\linewidth]{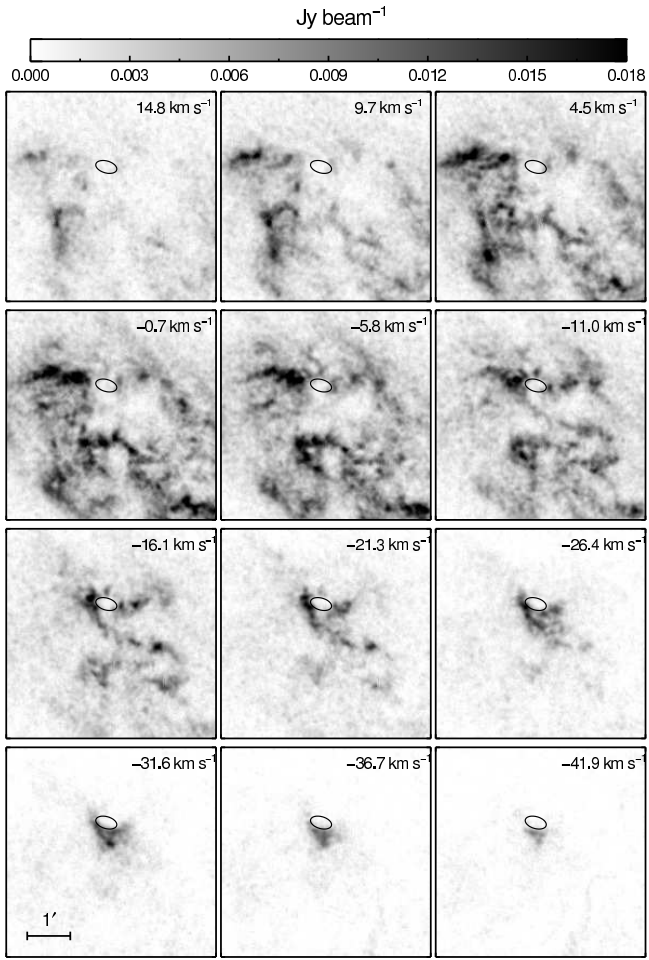}
\caption{The channel map of the SGS region in the `derotated' \HI data cube. The velocity of each channel is shown in the top right corner of the panel.
The black ellipse indicates the position of the hole inside the SGS.}\label{fig:HI-channel}
\end{figure}

\begin{figure}
\includegraphics[width=1\linewidth]{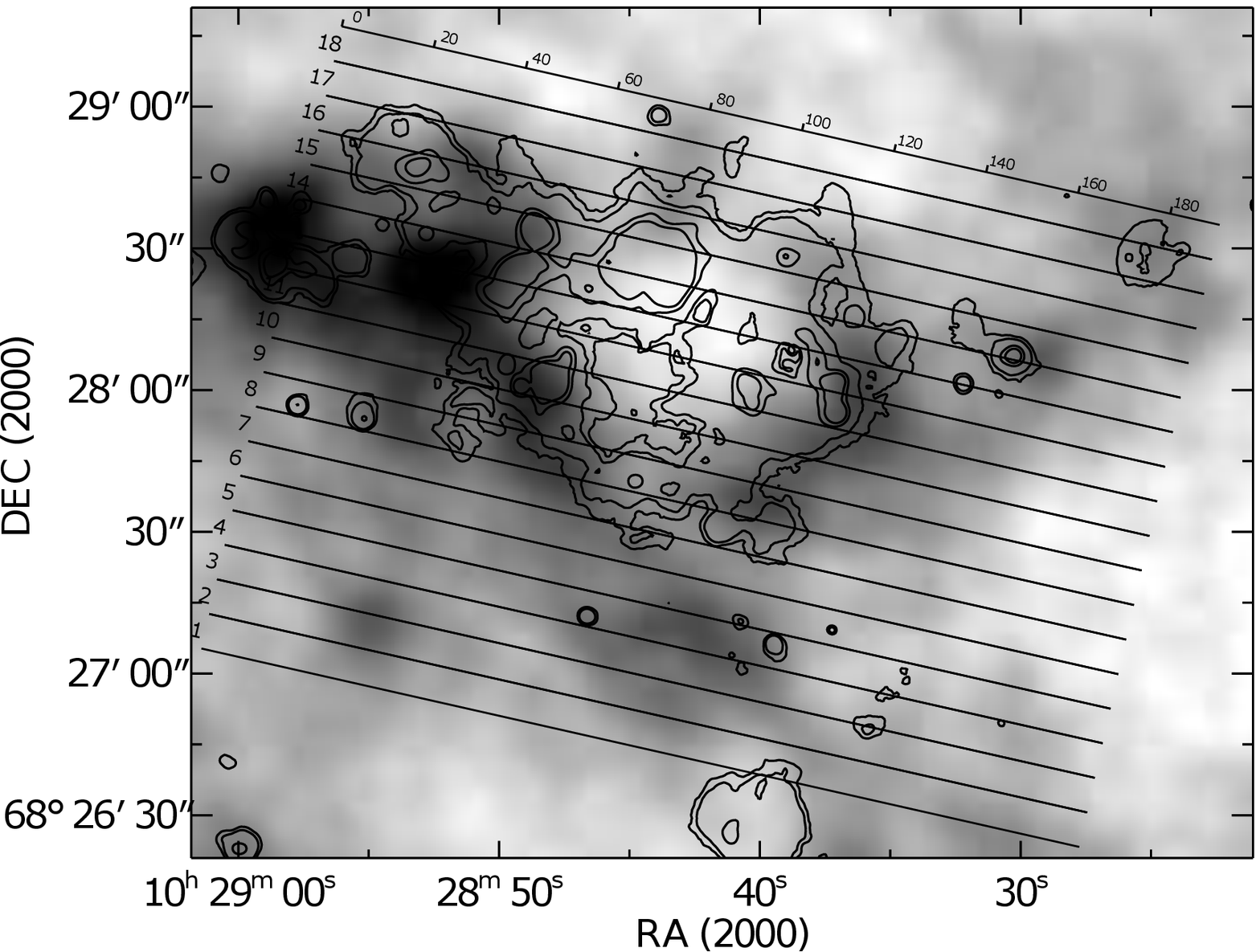}

\includegraphics[width=1\linewidth]{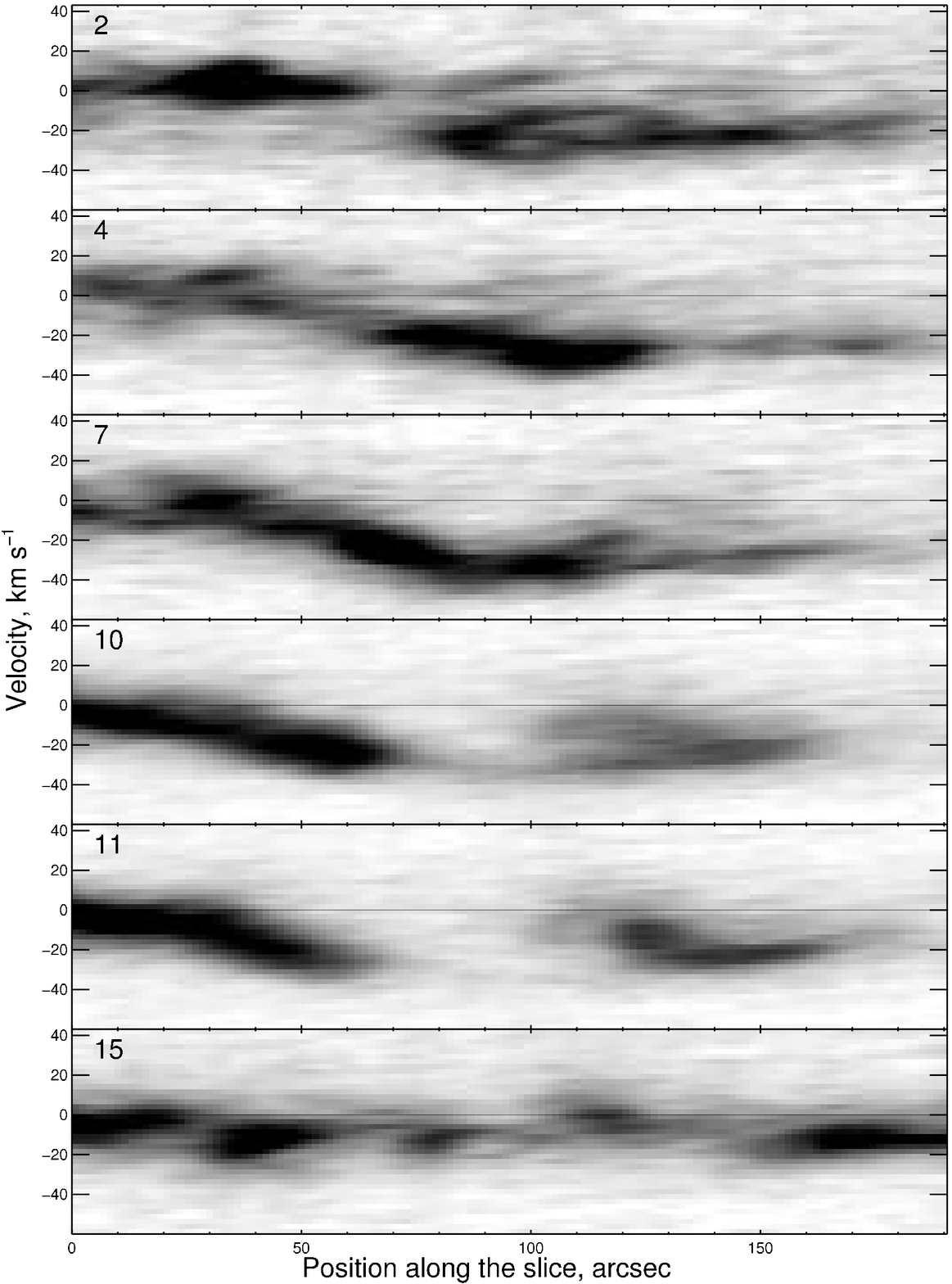}

\caption{Location of the PV diagrams on the \HI map of the SGS. The
contours correspond to the \Ha emission intensity levels of
$[0.5, 1.0, 5, 10]\times10^{-16} \mathrm{erg\ s^{-1} cm^{-2}}$.
Selected PV diagrams are shown in the bottom panels.}\label{fig:global-pv}
\end{figure}

We made two wide PV diagrams along the SGS major and minor axes (their full width was 75 and 45~arcsec, respectively).
Figure~\ref{fig:thick-pv} shows the location of these wide scans and the corresponding PV diagrams. A clear `velocity ellipse' corresponding to the emission of the approaching and receding sides of the shell is seen on the `derotated' PV diagram \#~1 in this figure.

To provide an accurate estimate of the expansion velocity of neutral gas in the entire SGS region we made a Gausssian decomposition of the \HI 21 cm line profile at several positions along the PV diagram. These positions are shown in Fig.~\ref{fig:thick-pv}, and the
corresponding line profiles -- in Fig.~\ref{fig:thick-prof}. All profiles show a
complex structure and can be decomposed into two or three components. The maximum velocity difference
between the components is observed in line profile \#~2 from the entire SGS region. This difference
is $48 \pm 4 \kms$, implying that the expansion velocity\footnote{Hereinafter, we give the formal uncertainties of the expansion velocity measurements obtained by fitting synthetic spectra with the same signal-to-noise ratio as the observed spectra} in this area is
about $24 \pm 2 \kms$.
Profile \#~4 constructed for the SGS region at the position shown in the bottom PV diagram
in Fig.~\ref{fig:thick-pv} is not separated into components, but its FWHM is the same as for
profile \#~2. All these results confirm the value $v_\mathrm{exp} = 25 \kms$
estimated  by \citet{walter99}.

The vicinity of the SGS is a place where several contacting and, possibly, interacting shells are located.  It is the only region
showing significant deviations of the line-of-sight velocity field from the fitted circular rotation model (see the bottom right panel in Fig.~\ref{fig:velmodel}).
If these peculiar velocities occur in the
disc plane, they should represent the inflow of gas into the large-scale stellar bar, because in the case of
a normal trailing spiral the SE-half of the galaxy's disc is nearest to the observer. An alternative
explanation of the observed kinematics is their attribution to vertical gas motions caused, for example,
by the fall of an \HI
gas cloud or vice versa, by SGS expansion. The channel map of the `derotated' data cube (Fig.~\ref{fig:HI-channel}) reveals that these negative residual velocities are related to the high-velocity ($-25 - -45 \kms$) \HI cloud, which moves towards us and is located south of and adjacent to the SGS. It is the only place in the galaxy that emits in the \HI 21 cm line at such high negative velocities. At smaller velocities the channel map shows a pattern typical of expanding shells: the bright approaching southeastern part of the SGS can be seen mostly at negative velocities, whereas the faint receding northwestern part shows up
at positive velocities. Note that the SGS has a non-uniform structure which is very different from the regular elliptical shape that most authors use to describe the
central hole. The observed elliptical shape of the SGS is due to the  inclination of the galaxy, $i=53\degr$  \citep{Oh08}. In the first approximation, the shell has a round shape if seen in the plane of the galaxy.
Because of the non-uniform structure
and kinematics of neutral gas in the extended NE region of the galaxy, we decided to analyse the \HI
kinematics  in the region in more  detail. For this aim, we created 18 PV diagrams along the SGS major axis, each  with a width of 7.5~arcsec; they cover a $3.2\times2.3$ arcmin area around the SGS.  Their location is shown in the top panel of Fig.~\ref{fig:global-pv}; the bottom panels show the most interesting PV diagrams. The approaching
side of the shell shows up conspicuously in these diagrams, whereas the contribution
of the receding side to the neutral hydrogen emission is barely discernible. This may be due to
the fact that the density of the ambient gas at the receding side of the SGS is lower than
at the approaching side. Hence the SGS must reside at the far side of the galactic disc.

\section{Ionized gas kinematics in star formation regions}\label{sec:HII}

\begin{figure}
\includegraphics[width=\linewidth]{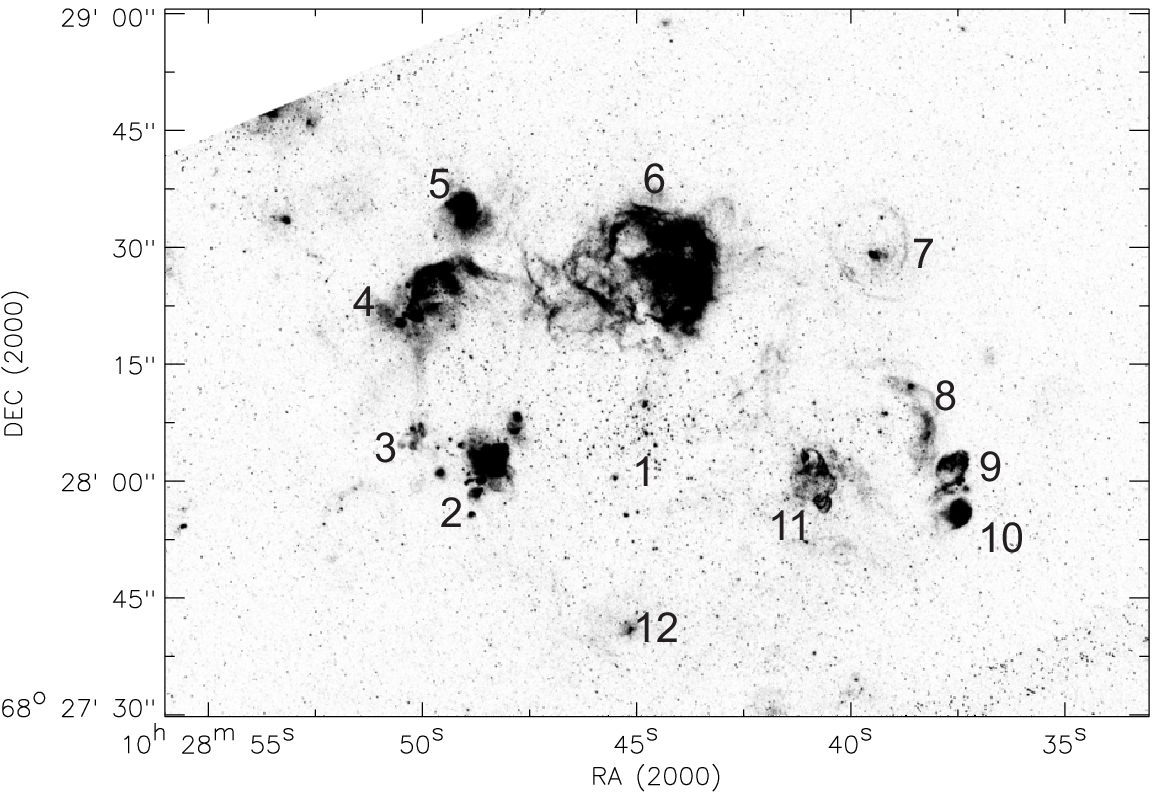}
\includegraphics[width=\linewidth]{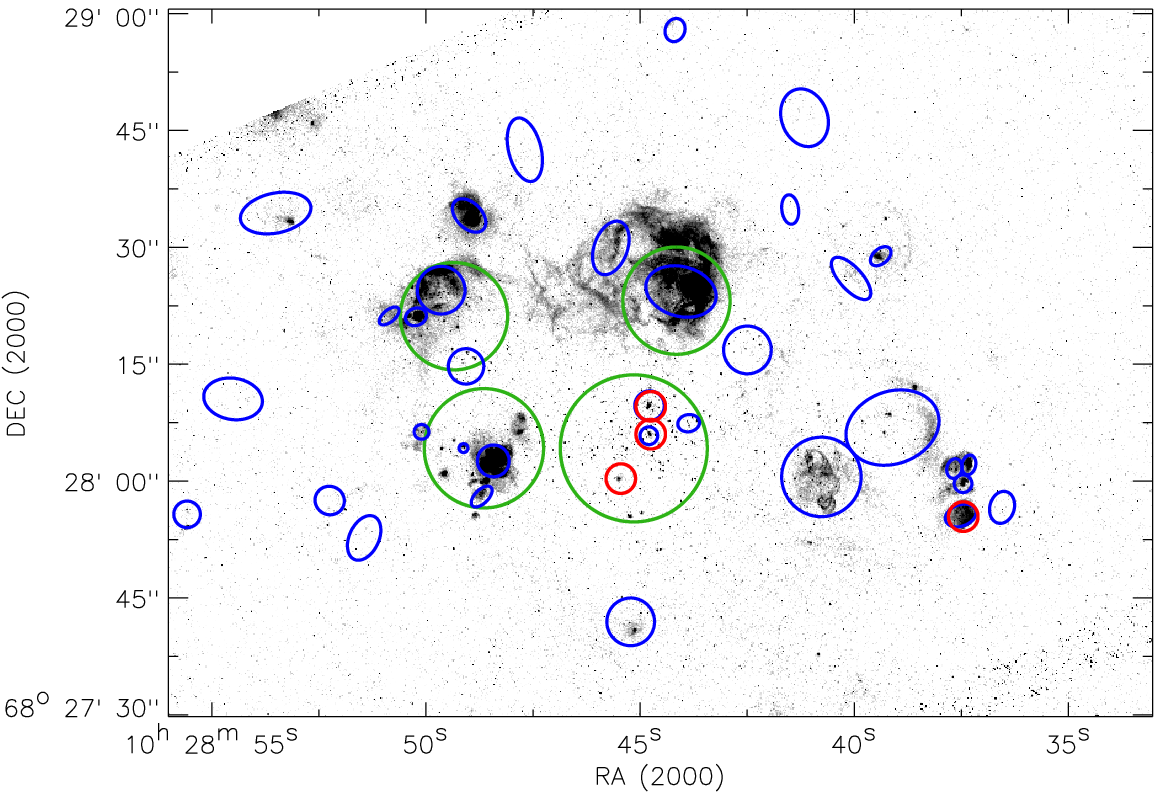}
\caption{Top: Identification of giant \HII regions on the \textit{HST}/ACS F658N image of the SGS
area according to the list of \citet{stewart00}.
 Bottom: location of the star clusters on the same image. The clusters identified by \citet{pellerin12} are shown by blue. The red circles denote the clusters from \citet{cook12}.
The four largest green circles show the star clusters identified by \citet{yukita12}.}
\label{fig:hii-regs}
\end{figure}

The rims of SGS contain numerous groups of triggered star-formation that are observed as   \HII complexes. The \Ha image of the region obtained with \textit{HST}/ACS is shown in Fig.~\ref{fig:hii-regs}. The complexes of star formation are denoted by numbers according to their identification in \citet{stewart00}. Note that these authors named the region of FUV emission around the central star cluster inside the SGS `Region \#~1'. They noticed the absence of \Ha emission in this region. We detected faint extended \Ha emission inside the interior of the SGS. We therefore designated as Region \#~1 not only the area near the central cluster, but the entire interior of the SGS.

 To study the nature of the \HII complexes  in the wall  of the SGS we
 evaluated their expansion velocities.
Given the size $R_{s}$ and expansion velocity $v_\mathrm{exp}$ of the \HII shell one can estimate its kinematic age $t$ and the energy input rate $L_{w}$ needed for its formation using the relations that describe the evolution of the
supershell driven by supernovae and stellar wind  \citep{maclow88}:

\begin{equation}
 R_{s}(t)= (125 L_{w}/154 \pi \rho _{o})^{1/5} t^{3/5} = 67 (L_{38}/n_{o})^{1/5}
t_{6}^{3/5}~\mathrm{pc}
\label{eq:rad}
\end{equation}
\begin{equation}
v_\mathrm{exp}(t)= 0.6 R_{s}/t = 39.4 (L_{38}/n_{o})^{1/5}
t_{6}^{-2/5}~\kms
\label{eq:vel}
\end{equation}

where $ L_{38}= L_{w}/10^{38}\ \mathrm{erg\ s^{-1}}$, $ t_{6}= t/10^{6}\ \mathrm{yr}$, and $n_{o}$ is the ambient gas volume density (in our case, the ambient \HI density $n_{\mathrm{HI}}$).
According to \citet{walter98}, the mean value in the SGS is $n_\mathrm{HI} = 2 - 3\ \mathrm{cm}^{-3}$. The density distribution in the SGS walls is highly non-uniform
and therefore for our analysis of \HII complexes we estimate the local unperturbed \HI density
in their neighbourhood.
We use two methods to convert the observed \HI surface density into \HI volume density.

\begin{table*}
\caption{Parameters of the star forming regions}\label{tab:hii_params}
\begin{threeparttable}
\begin{tabular}{lrcllllll}
\hline
\# of region & Size, pc  &\multicolumn{1}{c}{$v_\mathrm{exp}^*, \kms$}&$t$, Myr & \multicolumn{2}{c}{$n_\mathrm{HI}$, cm$^{-3}$}  & $L_{38}, 10^{38} \mathrm{erg\ s}^{-1}$  & Age, Myr    & E/Age, $10^{38} \mathrm{erg\ s}^{-1}$  \\
SW (YS)      & \textit{HST}  & \multicolumn{1}{c}{FPI}  & FPI  & [1] & [2] & FPI & SW (YS)   & SW (YS)      \\ \hline
(1) & (2) & (3) & (4) & (5) &(6) &(7) &(8) & (9)  \\ \hline
       1 (C2)&$800^{**}$ & $25^{**}$     & $9.6^{**}$   & $0.05$  & $0.3 $         &  $0.5 - 2.7^{**}$      & 8.5 (17) & 1.5 (29)   \\
       2 (C3)&$90\times150$   & $\le 19$  & $\ge1.9$& $0.63$  & $2.6 (2.93)$  & $\le(0.05 - 0.23)$      &  3.5 (17) & 0.8 (16)   \\
       3     & $80\times55$   & $\le 18$  & $\ge1.1$& $0.70$  & $3.0 (2.93)$   & $\le(0.02-0.06) $   &  4.0      & 0.24      \\
       4 (C4)&$150\times250$   & 27   & 2.4  & $ 0.56 $ & $ 2.5 (2.7) $   &$0.5-2.2$ &  4.2 (7)  & 0.9 (4.7)  \\
       5     &$65\times95$    & $\le18$  & $\ge1.5$& $0.53$  & $2.0 (2.7)$   & $\le(0.02-0.09)$      &  3.5      & 0.22     \\
       6 (C1)&$450\times320$&30   & 4.0  & $0.20$ & $0.9 (1.03)$& $0.8-3.5$   &  4.1 (7)  & 2.4 (27)   \\
       7     & 210  & 65   & 1.0  & $ 0.23 $ & $0.9 (3.02)$    & $2.5 - 9.9$   &  4.2      & 0.11      \\
       8     & 260  & 45   & 1.7  & $0.33 $ & $ 1.5 (2.54)$    & $1.9-8.4$  &  4.5      & 0.23      \\
       9     & 90   & $\le20$  & $\ge1.3$& $0.55 $  & $2.3 (2.54)$ &$\le(0.03 - 0.14)$   &  3.4      & 0.04    \\
       10    & 65   & $\le20$  & $\ge1.0$& $0.55 $ &$ 2.0 (2.54)$  &$\le(0.02 - 0.06)$   &  3.4      & 0.03     \\
       11 SNR& $110\times175$ &$70-80$ & $0.3^{***}$ & $0.20 $ & $0.8 (3.52)$      &see text&  4.3      & 0.32      \\
       12    & 40   &  $\le20$  & $\ge0.6$  & $ 0.40 $& $1.5 (2.75)$      &   $\le(0.005 - 0.017)$       &  4.1      & 0.07       \\
\hline
\end{tabular}

\begin{tablenotes}
\scriptsize
\item SW -- data derived by \citet{stewart00};
\item YS -- results obtained by \citet{yukita12} for complexes C1, C2, C3 and C4 (in parentheses);
\item \textit{HST} -- estimated based on the \textit{HST}/ACS F658N image;
\item FPI -- obtained from our FPI observations in the \Ha line;
\item * the typical uncertainty of $v_{\mathrm{exp}}$ is $2 \kms$;
\item ** estimates based on the  \HI data;
\item *** obtained using the \citet{sedov} solution for the SNR, see Section~\ref{sec:SNR}.

\end{tablenotes}

\end{threeparttable}
\end{table*}

The first method consists of the use of the relation between the column density $N_\mathrm{HI}$ and the volume density $n_\mathrm{HI}$: 
\begin{equation}
N_\mathrm{HI}=\int_{-\infty}^{+\infty}n_\mathrm{HI}\exp\left(\frac{-z^2}{2h^2}\right)dz=\sqrt{2\pi}hn_\mathrm{HI},
\end{equation}
where we adopt the scale height of the \HI disc of the galaxy $h=350$~pc  \citep{walter99}.
The inclination of the galactic disc increases the scale height along  the line of sight as $h/\cos i$.
We adopted  $i=53\degr$ from \citet{Oh08}. Note that this method may underestimate the density,
because the line-of-sight \HI distribution is non-uniform and can be dominated by the contribution
of small dense clouds.

The second method is a modification of  the one suggested by \citet{walter99}. These authors  define areas with an effective radius $r_\mathrm{eff}$ around each of the \HII regions and  assume that \HI emission comes mostly from the volume of the sphere of radius $r_\mathrm{eff}$. We believe it to be more correct to compute the volume  where the \HI emission emerges as the area of the cloud in the SGS wall multiplied by the wall thickness. We assume a uniform gas distribution and that the size of the gas layer in the wall is the same along the line of sight and in the sky plane. We found this size to be equal to 360~pc based on an analysis of several \HI clouds in the southern and western parts of the SGS. Unlike the first method, this one may overestimate the  volume density $n_\mathrm{HI}$, because it does not take into account the fact that \HI emission is integrated along the line of sight outside the volume considered.

 The  volume density estimates $n_\mathrm{HI}$ obtained for all the
\HII regions studied are listed in
Table~\ref{tab:hii_params}. The densities obtained using the first and second methods are marked by [1] and [2] in columns (5)  and (6),
respectively. The densities obtained by \citet{walter99} are given in parentheses.

The other columns of  Table~\ref{tab:hii_params} give: (1) the number of the complex according to \citet{stewart00} (shown in Fig.~\ref{fig:hii-regs}) and \citet{yukita12} (in parentheses);
(2) the size estimated from the \textit{HST}/ACS F658N image;
(3) $v_\mathrm{exp}$ obtained from our ionized gas kinematics analysis;
(4) the kinematic age calculated from equation (\ref{eq:vel});
(7) the  kinetic energy input rate required to drive the formation of the  complex computed from its size and expansion velocity;
(8) the age of the complex reported by \citet{stewart00} and \citet{yukita12} (in parentheses);
(9) the  kinetic energy input rate provided by the stellar population of the complex according to the
estimates of  \citet{stewart00} and \citet{yukita12} (in parentheses).

To reveal the effect of the expansion of  \HII complexes using the data of FPI observations in the  \Ha line we covered each of the \HII  regions in the SGS walls  by a series of  PV diagrams.  We use them together with the results of multi-component fitting of \Ha line profiles to estimate $v_\mathrm{exp}$, kinematic age $t$, and the required energy input rate $L_{38}$.
This way we measured the expansion velocities for 4 star formation complexes in the SGS and the upper limits of $v_\mathrm{exp}$ for the other regions.
  As an example of our analysis we will discuss in detail the largest Region \#~6. Figure~\ref{fig:reg6-pv} shows the \Ha PV diagrams and their location for this region. To avoid overloading the figures, for the other \HII complexes (except for the special case of Region \#~11) we marked in Fig.~\ref{fig:all-pv-loc} the locations of the most distinctive PV diagrams shown in Fig.~\ref{fig:all-pv}.
Examples of \Ha line profiles from all regions and the results of their Voigt decomposition are shown in Fig.~\ref{fig:all-prof}.

\subsection{\HII complexes with measured expansion velocities }


\begin{figure}
\includegraphics[width=\linewidth]{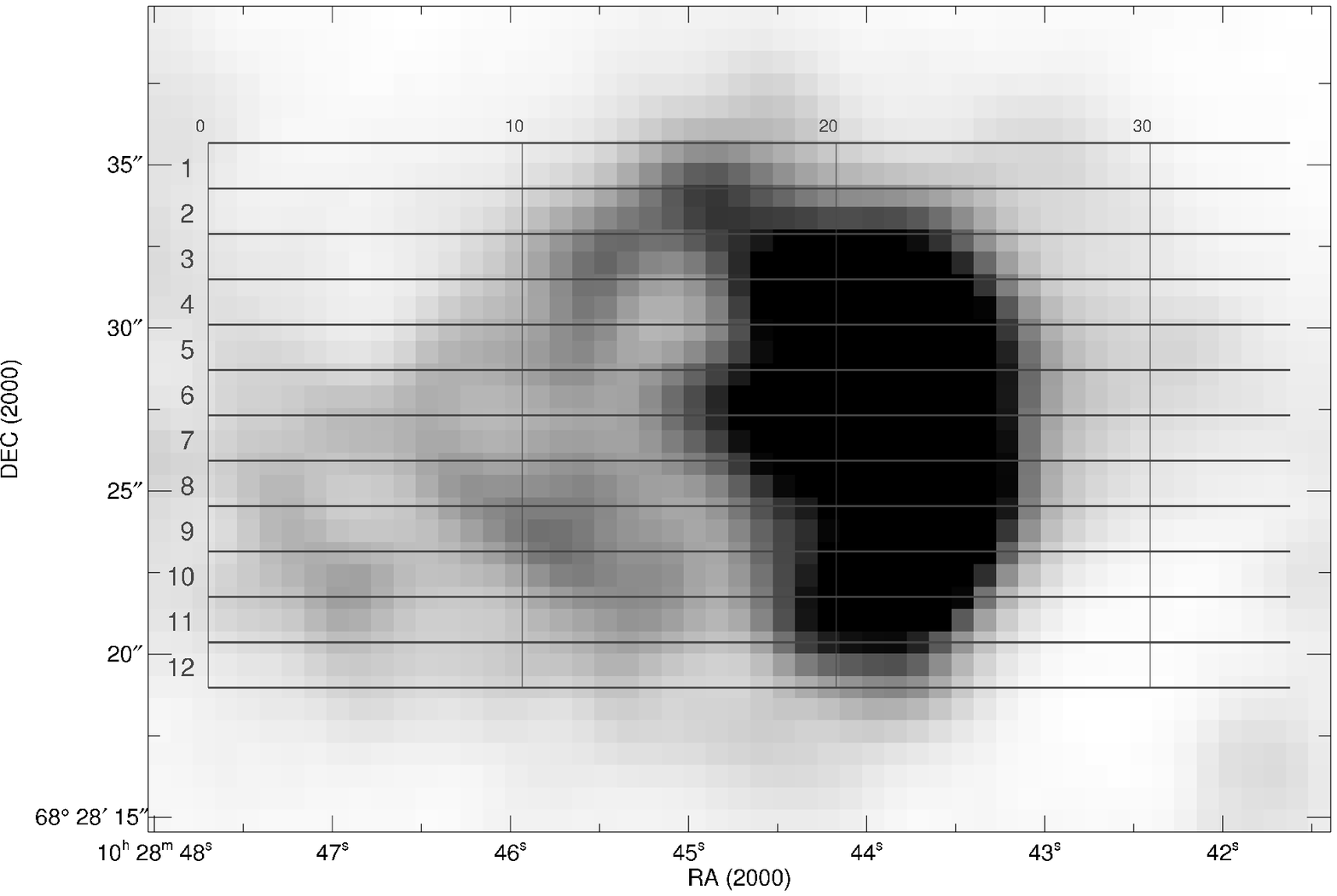}

\includegraphics[width=\linewidth]{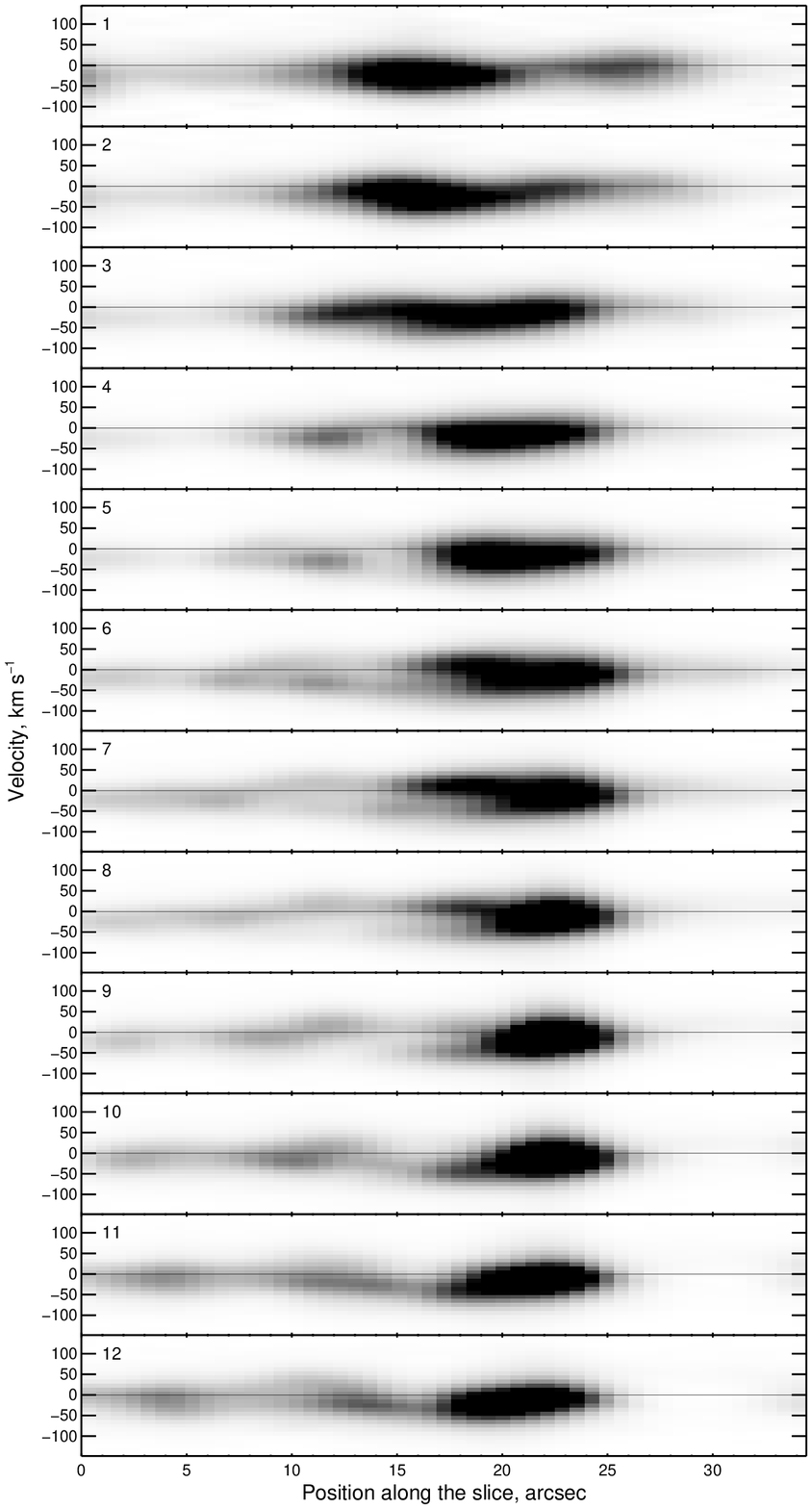}
\caption{Region \#~6: PV diagrams in $\mathrm{H}\alpha$ and their location on the \Ha image obtained from
the FPI data cube as a sum of the channel maps.}\label{fig:reg6-pv}
\end{figure}

{\bf Region \#~6~(C1).}
Star-forming complex \#~6~(C1) -- an extended ($\sim$~$450\times320$~pc) fine-filamentary \HII region --
is located in the northern wall of the  SGS.
\citet{yukita12} identified the young star cluster C1 inside Region \#~6.
 The C1 cluster includes PI 13z from the list of \citet{pellerin12}; cluster PI 07c is
located in the eastern part of the region (Fig.~\ref{fig:hii-regs}). \citet*{drissen93}
found three Wolf-Rayet star candidates  in Region \#~6.

The PV diagrams (Fig.~\ref{fig:reg6-pv}) reveal
several expanding shell-like structures inside the complex. In the brightest western part of the complex
the profile does not resolve into two components, but has a large  FWHM of up to $70 \kms$
compared to $\mathrm{FWHM} =  45-55 \kms$ in the neighbouring positions. The expansion effect shows up
conspicuously east of this bright region in the fainter filamentary part of the complex
(see the positions from 7 to 20~arcsec in Fig.~\ref{fig:reg6-pv}).
The corresponding profiles (6-1, 6-3, 6-5 in Fig.~\ref{fig:all-prof}) can be fitted confidently
by two components. The distance between the components is about $60 \kms$, while their
FWHM is $45-50 \kms$. These data yield a characteristic expansion velocity of $v_\mathrm{exp} \simeq 30 \kms$
for Region \#~6~(C1).

The eastern shell-like structure appeared at $0-7$~arcsec on the PV diagrams $7-12$
in Fig.~\ref{fig:reg6-pv} is probably not a part of  Region \#~6 but belongs to  Region \#~4, and the interaction of these two structures produces the faint details in  \Ha line profile 6-2
in Fig.~\ref{fig:all-prof}.

The average radius $R_s \simeq 200$~pc  combined with  $v_\mathrm{exp}=30 \kms$ yields $t = 4$~Myr and $L_{38}/n_{\mathrm{HI}} \simeq 3.9$ according to relations~(\ref{eq:rad}) and   (\ref{eq:vel}).

We estimate the ambient gas volume density $n_{\mathrm{HI}}$ for Region \#~6 to be $0.2\ \mathrm{cm^{-3}}$ and
$0.9\ \mathrm{cm^{-3}}$ using the first and second method, respectively. The corresponding energy input rates are $0.8\times 10^{38}\ \mathrm{erg\ s}^{-1}$  and  $3.5\times 10^{38}\ \mathrm{erg\ s}^{-1}$.

\citet{yukita12} determined the age $t=7$~Myr and the current mechanical energy produced by
the central star cluster C1, $L_{w}\simeq 27\times10^{38}\ \mathrm{erg\ s}^{-1}$; according to \citet{stewart00}, the age of the cluster is 4.1~Myr and the mechanical energy input rate is $L_{w}\simeq 2.5\times10^{38}\ \mathrm{erg\ s}^{-1}$, see
Table~\ref{tab:hii_params}.
As is evident from the above, this energy is definitely sufficient to drive the formation
of the shell-like complex given the inferred expansion velocity.


{\bf Region \#~4.}
This region has the form of a bright `half-shell' whose eastern part adjoins the dense wall of the
SGS (see Fig.~\ref{fig:SGS-color}), with a fainter filamentary inner part. The full size of the region is about $250\times150$~pc, and the accepted characteristic radius is $R_s=110$~pc. The faint filaments in the northwest extend out to the faint eastern filaments of
Region \#~6. This region hosts the clusters  PI 21z, 22z, and 23z
(see Fig.~\ref{fig:hii-regs}).
In the PV diagrams (Fig.~\ref{fig:all-pv}) the intensity of the line decreases and its width increases in the direction from the bright edge of the half-shell toward its inner parts. The characteristic velocity of the region is about  $-15 \kms$. Inside the region
the line profiles (Fig.~\ref{fig:all-prof}) reveal a two-component structure with the
peak velocities  equal to $25$ and $-30 \kms$, respectively, implying an expansion
velocity of $v_\mathrm{exp} \simeq 27 \kms$.

{\bf Region \#~7.}
This region has a well-defined  structure of a thin ring with a diameter of 210~pc surrounding the
cluster PI 01b, with the cluster P3 01d located in the southeastern part of the ring. Inside the ring
there is a small ($\sim$~40~pc) bright emission region coincident with the cluster.

The PV diagrams and the line profiles  show a well-defined three-component structure in the ring nebula. The peak velocities of the components are equal to $-80$, 0, and $50 \kms$.
The FWHM of the shifted components are equal to about  $60 \kms$ and that of the central component is equal
to $45 \kms$, which corresponds to the halfwidth of the single line observed outside the ring  (see profile `diff-2' in Fig.~\ref{fig:all-prof}).
This  leads us to conclude that the central component is associated with
the emission of the
background gas in this region, and the shifted components -- with the approaching and receding sides of the expanding
shell. The corresponding expansion velocity is $v_\mathrm{exp} = 65~\kms$. The inferred expansion velocity implies
an age of $t = 1.0$~Myr for the ring-shaped \HII region, which is substantially smaller than the ages
of other star-forming complexes.

We discuss the possible nature of this unique ring nebula in
Section~\ref{sec:HII-results}.

{\bf Region \#~8.}
In the  \textit{HST}/ACS image (Fig.~\ref{fig:hii-regs}) a very faint broad ring can be seen
whose western part is Region \#~8 itself and the eastern part reaches Region \#~11.
The curvature of the eastern region indicates that these two parts may be considered as a single ring structure.
Region \#~8 itself appears as an arc with a chord length and curvature radius of about $220$~pc
and $130$~pc, respectively. Its bright western edge has a single-component
line profile (see profile \#8-1 in
Fig.~\ref{fig:all-prof}) and the line profile of the inner part has a two-component structure
(\#8-2 in  Fig.~\ref{fig:all-prof}). The velocity of one component varies little and is equal to
about $30 \kms$, whereas that of the other component varies from $-45$ in the east to $-65 \kms$
at the centre, implying an expansion velocity of about  $45 \kms$ at the centre.
The halfwidth of all the line components is  $40-45 \kms$.
In the PV diagrams shown in Fig.~\ref{fig:all-pv} Region \#~11 borders with
Region  \#~8 in its left part. As is evident from these diagrams, the extended ring expands as a whole.
The radius $R_s=130$ pc and expansion velocity $v_{\mathrm{exp}}=45 \kms$ of this  `combined' region imply
an age of $t=1.7$~Myr and a required energy input rate of
$(1.9-8.4)\times10^{38}\ \mathrm{erg\ s}^{-1}$.

\subsection{HII-complexes with estimates of the upper limit of the expansion velocity}

\begin{figure}
\includegraphics[width=\linewidth]{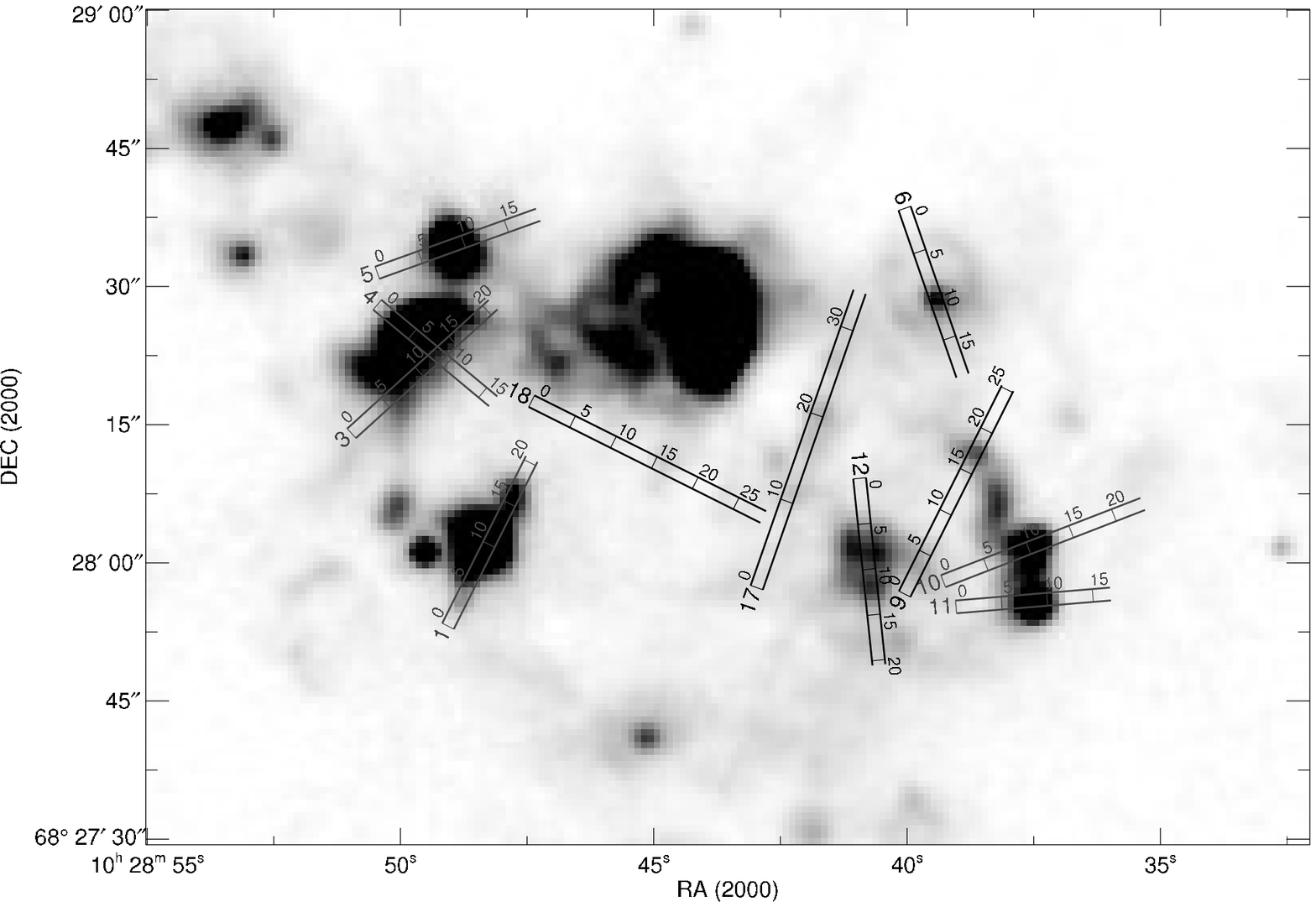}

\includegraphics[width=\linewidth]{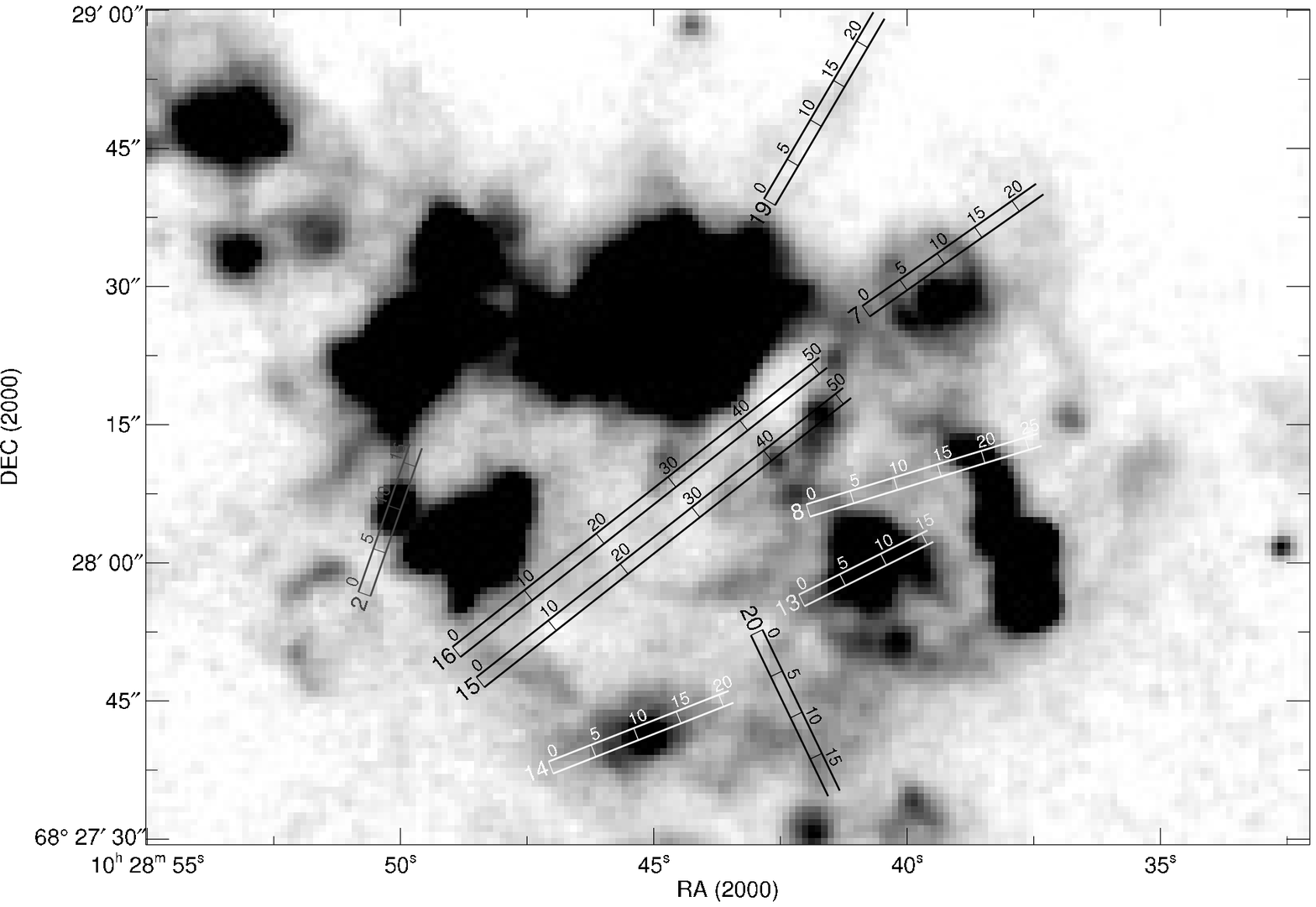}
\caption{The location of our PV diagrams that are shown as an example in Fig.~\ref{fig:all-pv}.
The images are the sum of the channel maps of the \Ha FPI data cube; they are shown at different intensity levels.}\label{fig:all-pv-loc}
\end{figure}

\begin{figure*}
\includegraphics[width=0.48\linewidth]{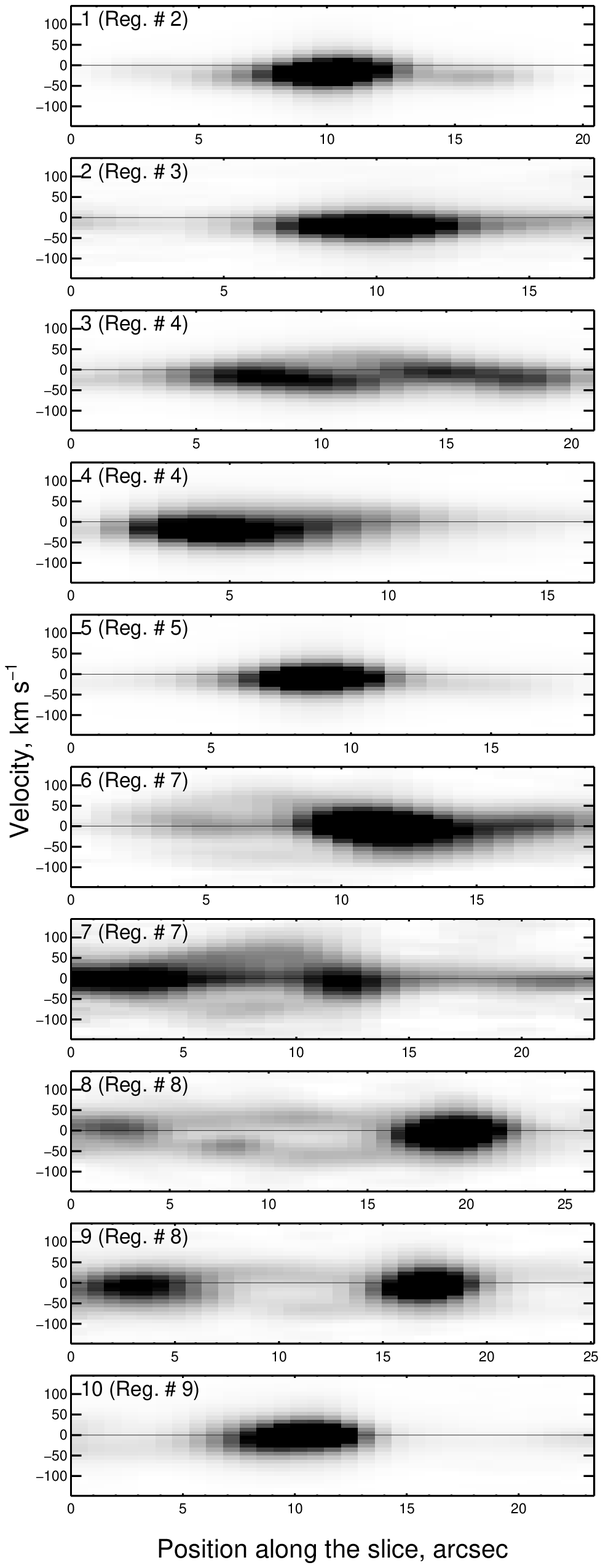}
\hspace{0.3cm}
\includegraphics[width=0.48\linewidth]{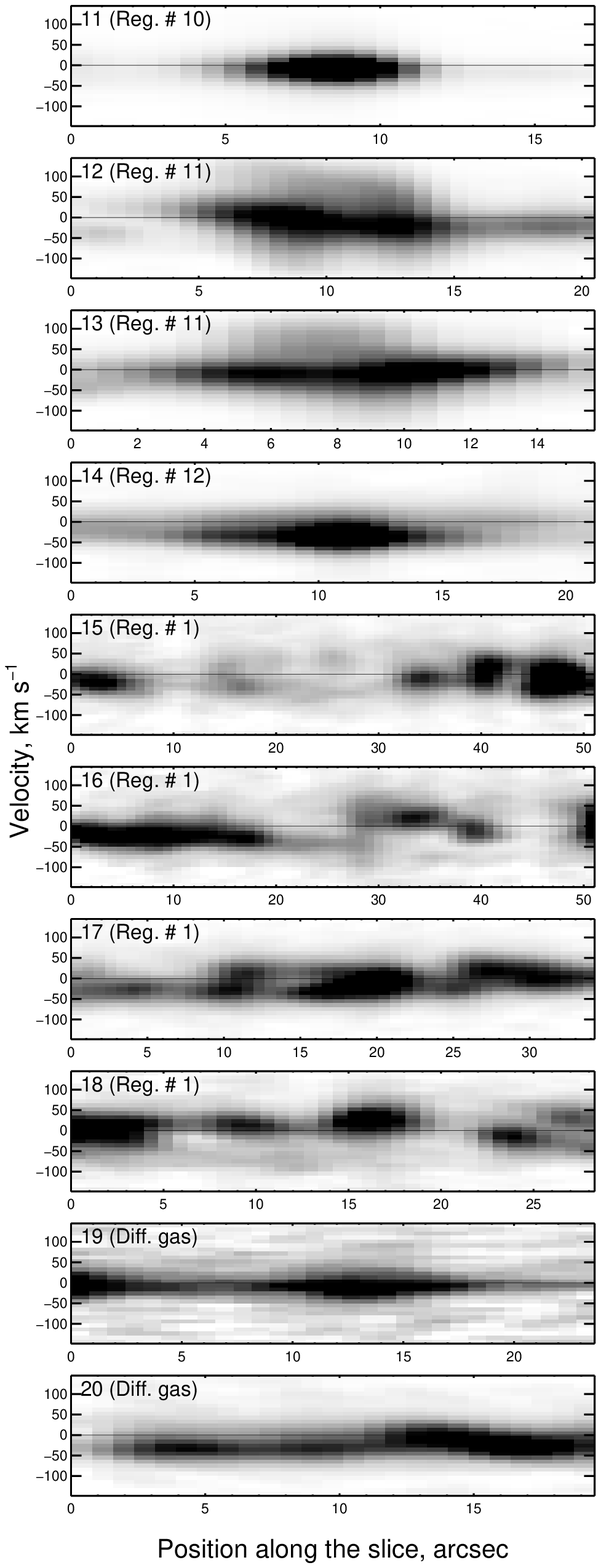}
\caption{Examples of PV diagrams in the  \Ha  passing through the observed \HII regions. Their
locations are shown in Fig.~\ref{fig:all-pv-loc}.}\label{fig:all-pv}
\end{figure*}

\begin{figure*}
\includegraphics[width=\linewidth]{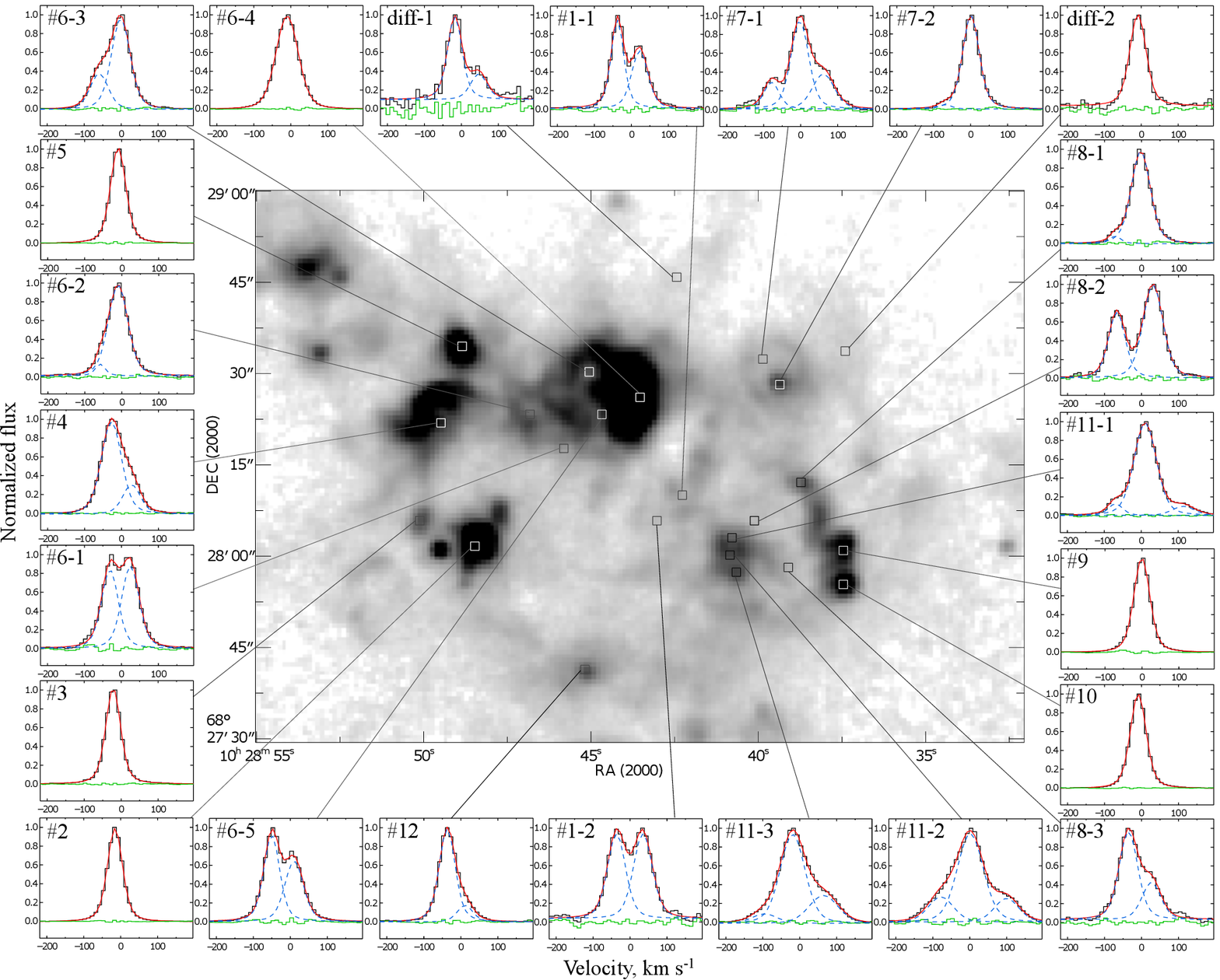}
\caption{Examples of \Ha line profiles obtained with the FPI. The black solid line denotes the observed data, the red solid line -- the fitted profile, the blue dashed line -- single Voigt components, and the green dashed line -- the residuals.}\label{fig:all-prof}
\end{figure*}

{\bf Region \#~2}, the brightest nebula in the SGS, is rather non-uniform and consists of a central nebula $\simeq 120$~pc wide, several more compact ($\simeq 10-20$~pc) clumps that surround it, and two
faint external shells southeast and northwest of the region with
the sizes of $\simeq 20$~pc and 35~pc (see Fig.~\ref{fig:hii-regs}).  The PV diagrams passing through the central nebula (see PV diagram \#~1 in Fig.~\ref{fig:all-pv}) show no clear signs of expansion.
The velocity of the peak of the single-component line profile (Fig.~\ref{fig:all-prof}) varies slightly over the $-15 - -25 \kms$ interval; the FWHM of the line profile corrected for the instrumental contour is $\simeq 38 \kms$.
We can estimate only the upper limit for the expansion velocity of the region as half of its FWHM,  $v_\mathrm{exp} \le 19 \kms$.
Hence the kinematic age of the complex is  $t \ge 1.9$ Myr and given the inferred density of the ambient
neutral gas  $n_\mathrm{HI}=0.6\ \mathrm{cm^{-3}}$ (method~1) and
$n_\mathrm{HI}=2.6\ \mathrm{cm^{-3}}$ (method~2) the required  mechanical energy input rate
does not exceed  $0.05\times 10^{38}$ and  $0.23\times 10^{38}\ \mathrm{erg\ s}^{-1}$,
respectively.

{\bf Region \#~3}, a faint $80\times55$~pc diffuse region, consists of several smaller ($\simeq 20$~pc) nebulae (see Fig.~\ref{fig:hii-regs}).  The PV diagrams and the line profiles show no
expansion effect; the line halfwidth  and peak velocity are equal to
$\mathrm{FWHM} \simeq 35 \kms$ and $v \simeq -25 \kms$, respectively.

{\bf Region \#~5}  consists of a relatively bright $40-50$~pc  compact core  located in the southern
part and a fainter, irregularly structured extended region about 70~pc to the north. The region hosts
the PI 24z cluster at its centre (see Fig.~\ref{fig:hii-regs}).
The PV diagrams  (Fig.~\ref{fig:all-pv}) show no noticeable features; the line profile
 has a single-component structure with an FWHM of about $38 \kms$; the mean line peak velocity is equal to $-10 \kms$.

{\bf Region \#~9} is a lenticular shell with a size of about 90~pc.  The line profiles   do not resolve into
two components and the  PV diagram  shows no clear signs of expansion. The average velocity of the region
is   about $-5 \kms$. The FWHM of the line is $40 \kms$ and $55 \kms$ at the centre and
periphery of the shell, respectively. The region hosts the clusters PI 25e, 25g, and PII 25h.

{\bf Region \#~10} has a round shape and a size of about 65~pc. It shows no signs of expansion either in the
PV diagrams or in the line profiles; the peak velocity and FWHM of the line are equal to about $-10 \kms$
and  $40 \kms$, respectively. The region hosts the cluster PI 25c.

\begin{figure}
\includegraphics[width=0.40\linewidth, angle=90]{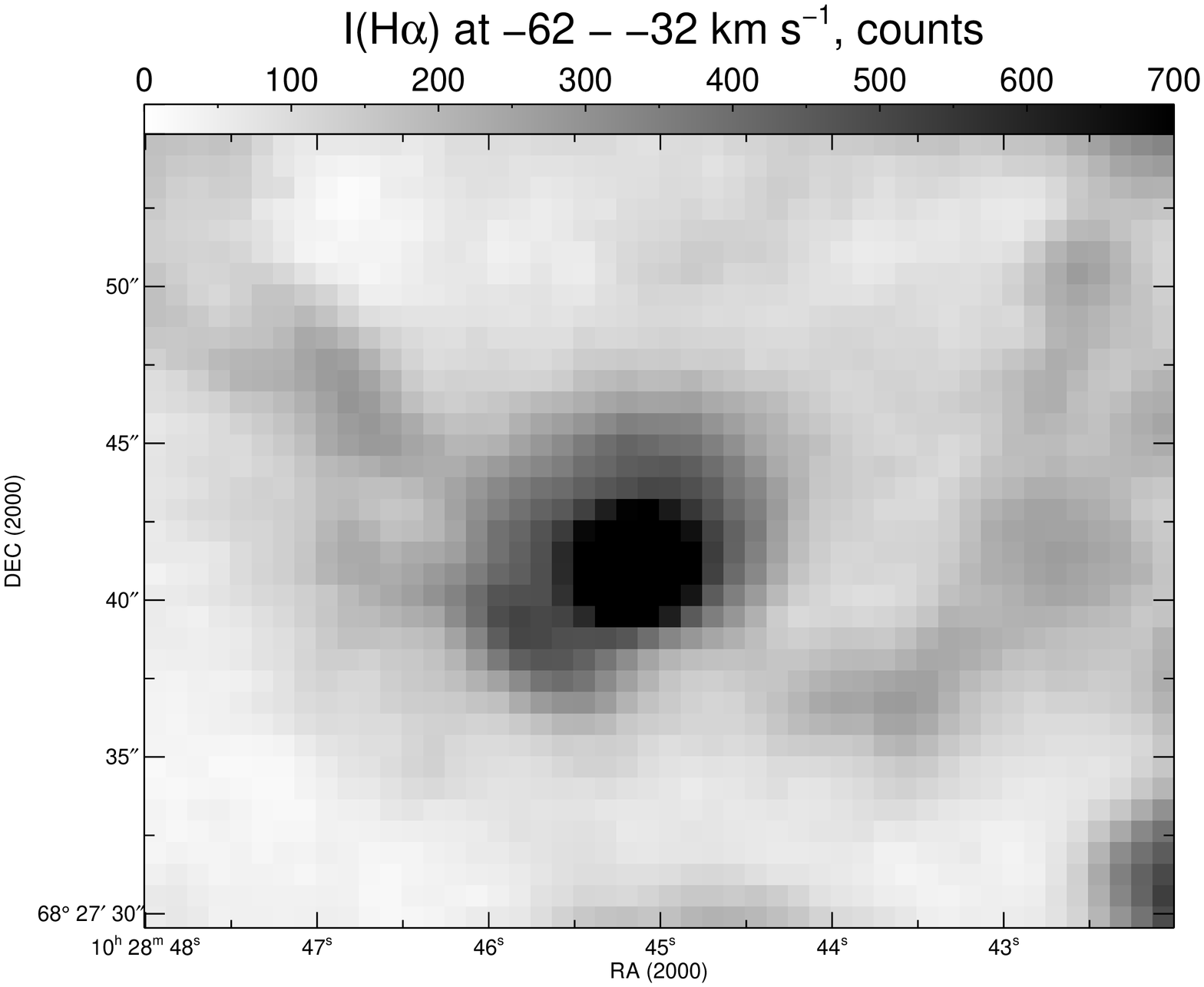}~\includegraphics[width=0.40\linewidth, angle=90]{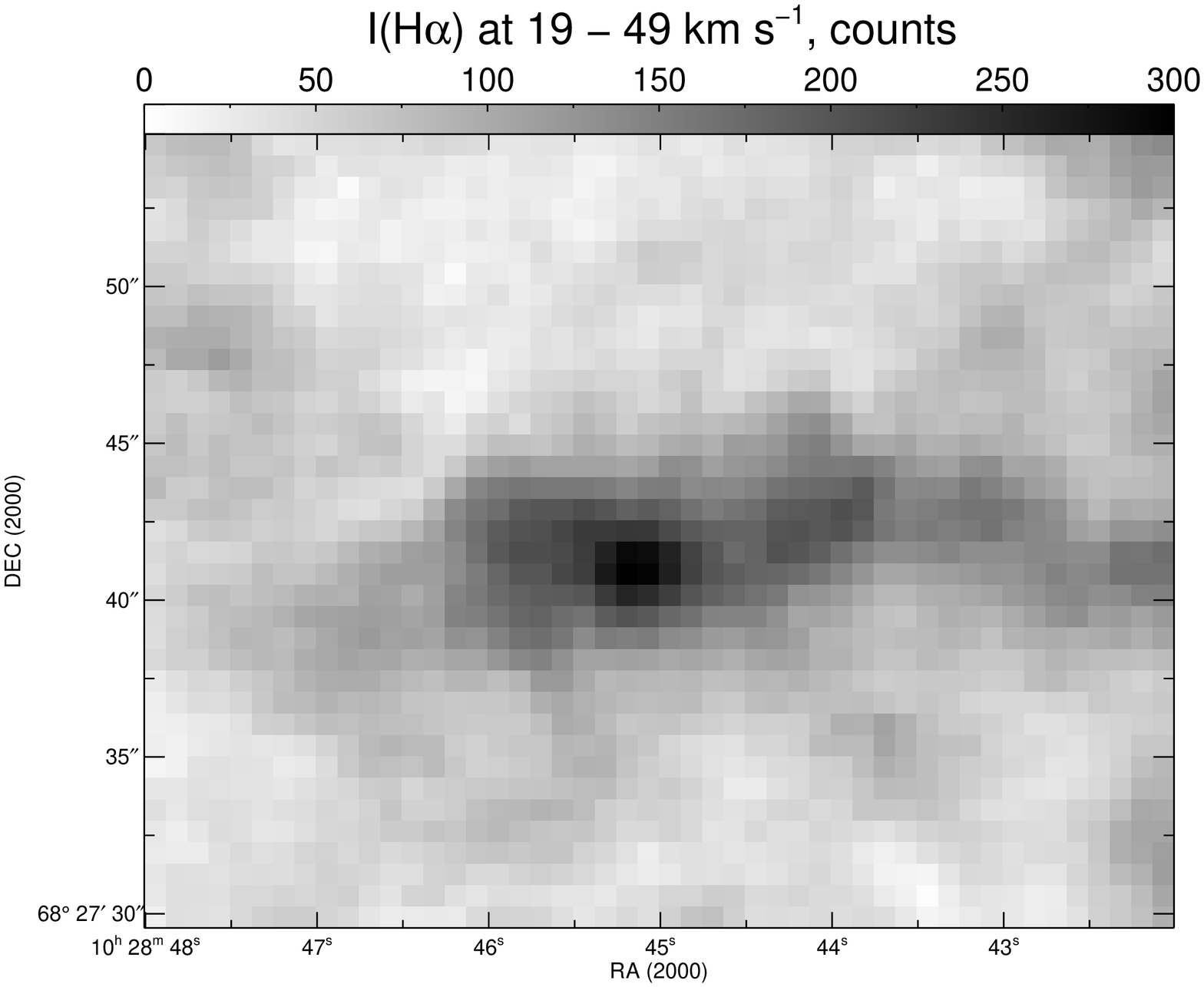}

\caption{Region \#~12: \Ha emission line maps at velocities ranging from $-62$ to $-32 \kms$ (left) and from $19$ to $47 \kms$ (right) based on the FPI observations.}\label{fig:reg12-chan}
\end{figure}

{\bf Region \#~12} has a  compact structure with a size of about 40~pc and a small bright core at its centre. The
region hosts the cluster PI 28b. The \Ha profile is asymmetric throughout the entire region
(Fig.~\ref{fig:all-prof}). A bright component is observed at a velocity of about
$-35 \kms$, and a second, weaker component, at $5 \kms$. The FWHM of both components is
$\sim40 \kms$. The shape of this region in the \Ha images at different velocities (Fig.~\ref{fig:reg12-chan}) reveals  two regions -- a bright compact one corresponding to the first profile and a faint extended region responsible for the second profile. Such a complex structure may be due to the fact that the object is located in the non-uniform region at the northeastern boundary of the high-velocity \HI cloud mentioned in Section~\ref{sec:HI}. That is why for this region we list  only the limiting values of  $v_\mathrm{exp}$,  $t$ and $L_{38}$ in Table~\ref{tab:hii_params}.

\subsection{Region \#~11 - an old supernova remnant.}\label{sec:SNR}

VLA radio observations  by \citet{walter98} have shown that the three brightest optical
star-forming regions in the SGS are the main sources of  thermal radio continuum emission (see their fig.1).
The fourth source located in the area of shell \#~11 is non-thermal and its radio emission is much
stronger than one would expect from an \HII region with rather weak \Ha emission.
The authors concluded that it is most likely a supernova remnant.

\begin{figure}
\includegraphics[width=0.74\linewidth, angle=90]{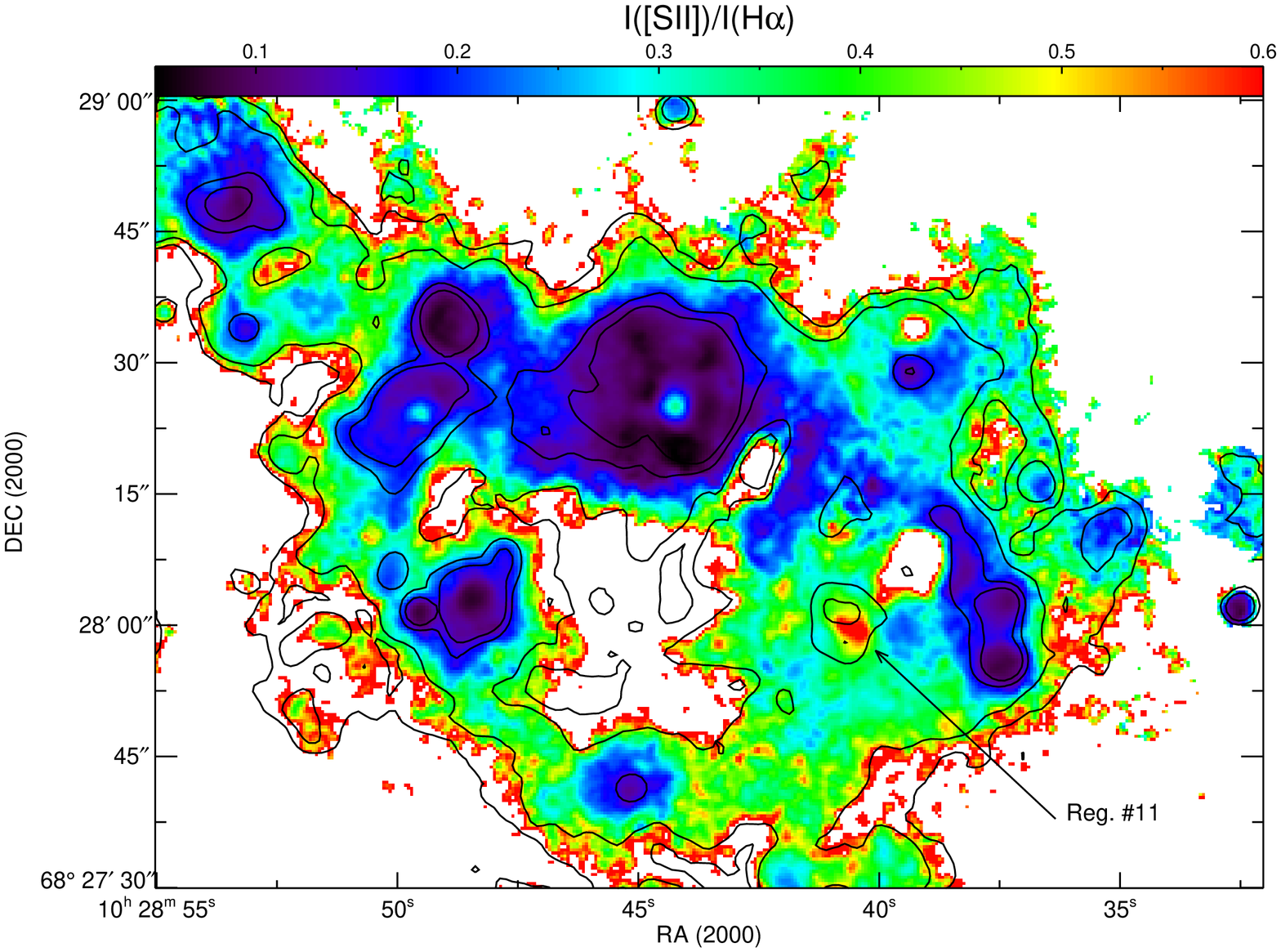}

\caption{Map of the \SIIHa\, flux ratio with the
$[0.5, 1.0, 5, 10]\times10^{-16} \mathrm{erg\ s^{-1} cm^{-2}}$ \Ha intensity
contours overlaid. The arrow indicates the position of Region \#~11 (SNR).}\label{fig:siiha}
\end{figure}

To verify this conclusion, we map the distribution of the  \SIIHa\ line intensity ratio based on the
SCORPIO narrow band direct images  (see Fig.~\ref{fig:siiha}).
The strong \SII 6717, 6731 \AA\ feature in the SNR area (\SIIHa\ $= 0.36 - 0.67$) is typical for cooling gas emission behind a shock front, which confirms the SNR identification \citep*[][and references therein]{allen08, millar12}.

\begin{figure}
\includegraphics[width=0.98\linewidth]{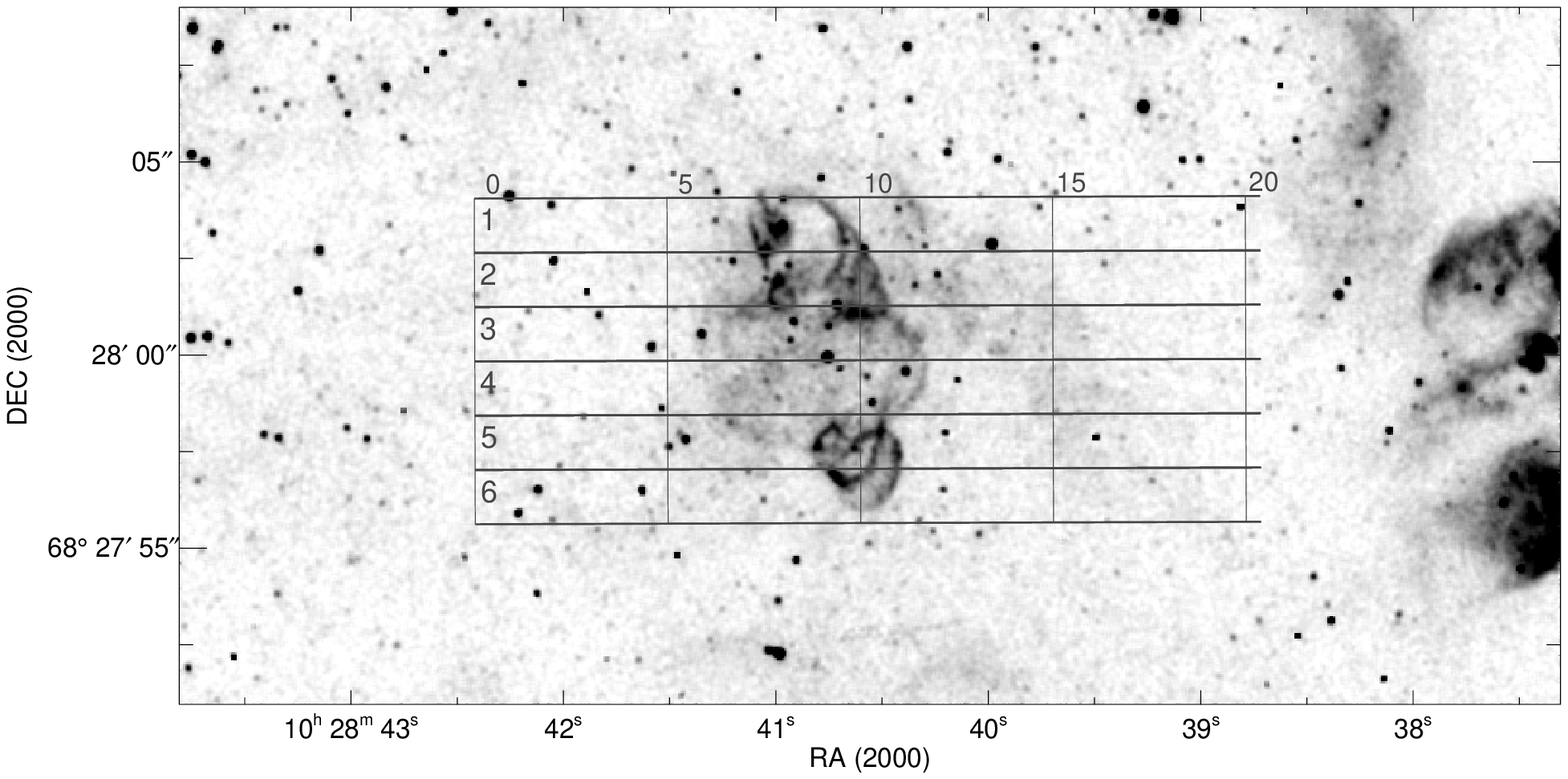}

\includegraphics[width=\linewidth]{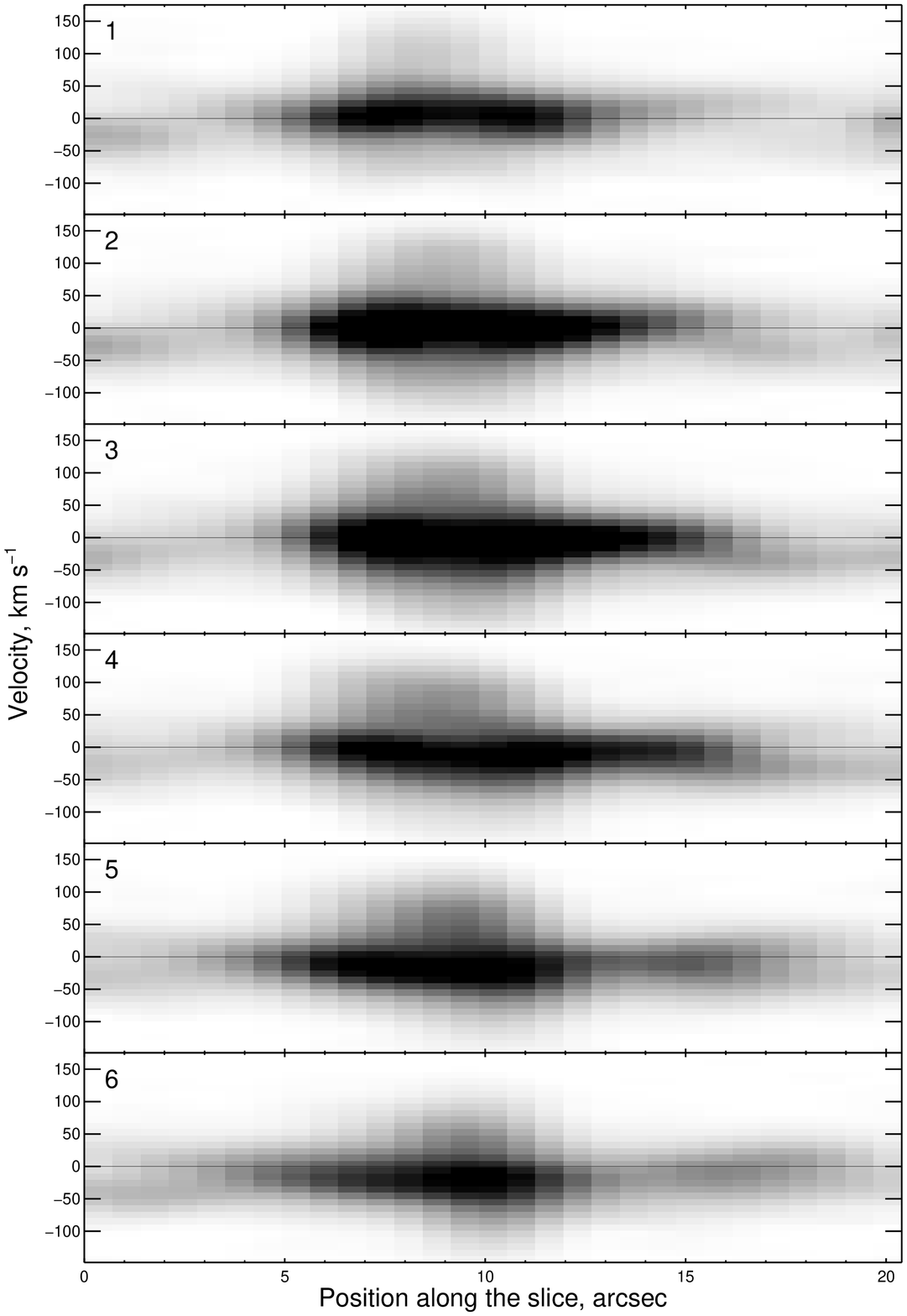}
\caption{PV diagrams in \Ha passing through Region \#~11 and their location on the
\textit{HST}/ACS F658N image of the region.}\label{fig:reg11-pv}
\end{figure}

The \textit{HST}/ACS image of the region (see Fig.~\ref{fig:reg11-pv}) reveals its filamentary
central shell ($D_\mathrm{horiz}=5.7\ \mathrm{arcsec} \simeq 110$~pc) and two $\sim40$~pc `lobes' north and south
of it (the full size is $D_\mathrm{vert} = 9\ \mathrm{arcsec}=175$~pc).
The region also hosts the  PI 02z  cluster (see Fig.~\ref{fig:hii-regs}).

The supernova remnant is characterized by high-velocity motions. The PV diagrams based on our FPI observations (Fig.~\ref{fig:reg11-pv}) reveal the presence of high-velocity gas in the SNR; the \Ha line profiles
(\#11-1, 11-2, 11-3 in Fig.~\ref{fig:all-prof}) show a well-defined three-component structure of
the line profile in the central part of the SNR.

All three components of the line profile are better seen in the northern part, whereas the negative
velocity component disappears towards the south. The central component has a velocity of about $-30 \kms$.
The FWHM of the central component is about $60 \kms$, and that of the shifted components
amounts to  $80-90 \kms$. The central component of the line  probably corresponds to the background emission of gas in the SNR region.

The peak velocities of the shifted components are about  $60-70 \kms$ and $-80 \kms$
and yield an expansion velocity of $ v_\mathrm{exp} \simeq 70 - 80 \kms$. This high expansion velocity provides further support for the identification of the nebula as an SNR.

The high expansion velocity suggests that the  SNR still undergoes  adiabatic expansion.
According to the self-similar solution of \citet{sedov}, the evolution of an SNR at the adiabatic
stage can be described by the following equations:

\begin{equation}
R_{s}=13.5(E_{51}/n_{o})^{0.2} (t/10^{4} \mathrm{yr})^{0.4} \ (\mathrm{pc})
\label{eq:rad-sn}
\end{equation}

\begin{equation}
v_\mathrm{exp}=0.4 R_{s}/t,
\label{eq:vel-sn}
\end{equation}
where $E_{51} = E_{o}/10^{51}\ \mathrm{erg}$ is the explosion energy.

Given the radius  $R_s=55$~pc  and the expansion velocity  $v_\mathrm{exp}=70-80 \kms$, we infer the SNR age of $t=(2.8-3.1)\times10^5$~yr and   $E_{51}/n_{o}=1.1-1.5$.
Our estimates of the unperturbed ambient density  $0.2\,\mbox{cm}^{-3}$ (method 1) and $0.8\,\mbox{cm}^{-3}$ (method 2) yield an initial explosion energy of $E_{o} = (0.2 - 0.3)\times10^{51}$~erg and
$E_{o} = (0.9 - 1.2)\times10^{51}$~erg, respectively.

If the SNR currently undergoes the post-adiabatic snow-plough stage of evolution, we can use the
analysis performed by \citet{chevalier74}. According to this work,

\begin{equation}
E_{50} = 5.3\times 10^{-7} n_{o}^{1.12} v_{\mathrm{exp}}^{1.4}(\kms) R_s^{3.12}(\mathrm{pc}),
\end{equation}
where $E_{50} = E_{o}/10^{50}\ \mathrm{erg}$.

In this case  the same SNR radius and expansion velocity yield the $E_{51}/n_{o}^{1.12}$ ratio
of $5.5 - 6.6$; the two ambient density estimates
yield $E_{o} \simeq 1.0\times 10^{51}$
and $E_{o} \simeq 4.7\times 10^{51}$~erg, respectively.

Thus our analysis of SNR kinematics confirms that Region \#~11 can be viewed as
an old remnant of a standard SN explosion.

We can speculate that the \Ha morphology of the SNR in IC~2574 considered here
(a central shell and two symmetrical  `lobes' north and south of it) resembles very much
the structure of the well-known Galactic SNR W50 with two symmetric eastern and western  `lobes'
due to jets emerging from the central  SS~433 microquasar \citep*[see][for a recent analysis] {goodall11}.


There is a noticeable asymmetry in the extent of the northern and southern lobes of the SNR in
IC~2574: ($R(\mathrm{N})/R(\mathrm{S})\simeq 1.35$), which is similar to the asymmetry  of the eastern
and western lobes of W50  ($R(\mathrm{E})/R(\mathrm{W}) =1.4$).
According to \citet{goodall11}, the  asymmetry of W50 is due to the exponential density profile of
the Milky Way disc. We can similarly explain  the  asymmetry in the extent of
the lobes of the SNR in IC~2574 by the density gradient in the wall of the SGS.

A discovery of an X-ray binary in the SNR could provide evidence for the similar nature of the two
objects.  Chandra  observations of the SGS by \citet{yukita12}
reveal a luminous, $L_{X} \simeq 6.5\times 10^{38}\ \mathrm{erg~s}^{-1}$ (in the $0.3-8.0$~keV band),
point-like source inside the hole of the SGS, but the separation between the source and the SNR is about
15~arcsec.

 To sum up, we can confidently conclude that Region \#11 really is an old SNR.

\section{Discussion}\label{sec:HII-results}

\subsection{The expansion of the \HII shells and their energy budget}\label{sec:regEnergy}

We have measured bona fide expansion velocities of four \HII complexes (\# 4, 6, 7, and 8) from the splitting
of the line profile. We can therefore use the above equations
(\ref{eq:rad}, \ref{eq:vel}) for the supershell produced by supernovae and stellar winds  to estimate the kinematic ages and mechanical energy
input rates  required to drive the formation of each  \HII complex. For the remaining  \HII regions we estimated only the upper limits for the
expansion velocity and
the corresponding lower age limit and upper  $L_{w}$ limit.

\citet{stewart00} report the results of FUV and optical observations of the SGS, providing an
independent measure of the ages, masses and total mechanical energy deposited by the central stellar
association and  by the clusters in giant \HII regions on the rim of the SGS based on the FUV, B-band,
and \Ha fluxes (see table~2 in their paper).

\citet{yukita12} analysed Chandra X-ray observations of the SGS. The authors  performed a multi-wavelength
(\textit{GALEX} and \textit{Spitzer}) analysis of the stellar population in
the SGS area and identified four star forming regions - young star clusters inside and at the edges of the
SGS (C1, C2, C3, C4 in their table~1 and fig.~2); C2 coincides with the central star forming region
studied by \citet{dalcanton12}. Their positions  are indicated  in Fig.~\ref{fig:hii-regs}
by the large green circles. These   largest clusters include the small stellar groups
from \citet{pellerin12},
see Fig.~\ref{fig:hii-regs}. The above authors determined the mass, age,
and mechanical energy produced by massive stars and supernovae in these four massive star clusters.
They estimate the rate of mechanical energy input from the X-ray fluxes in accordance with
the best-fitting \textsc{starburst99} model.

When comparing   our study of \HII regions  with the analysis
of their stellar populations one must  bear in mind the uncertainty of our velocity estimates on
the one hand and the fact that the reported parameters of star clusters are model dependent on the other
hand. The uncertainty of our estimates is due primarily to the non-uniform density of neutral gas in the
walls of the SGS and the irregular morphology and kinematics of \HII complexes.

However, Table~\ref{tab:hii_params} shows   that the mechanical energy
input rate for the overwhelming majority of the
complexes of triggered star formation associated with the SGS  is sufficient to drive the formation of
expanding shell-like  \HII regions. The only
exceptions are regions \#~7 and \#~8.

The fact that our inferred  energy input rate required to drive the formation of Region \#~8 significantly
exceeds the energy input rate estimated by  \citet{stewart00} is quite understandable. We
significantly increased the size of this  \HII region (and consequently the required energy) by incorporating into it the faint extended
arc-shaped nebula that adjoins the SNR from the west, i.e., by
combining two faint nebulae into a single shell-like structure.

\begin{figure}
\includegraphics[width=\linewidth]{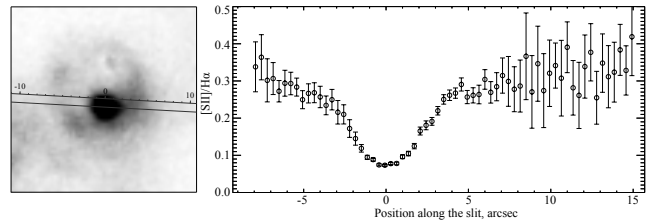}

\caption{The position of the slit crossing Region \#~7 in the \Ha image mapped from our FPI observations (left) and the
\SII 6717,6731 \AA\ to \Ha flux ratio along the slit (right).}\label{fig:reg7-ls}

\end{figure}

The situation with Region \#~7 is very interesting. We estimate $v_\mathrm{exp}\approx65\kms$ and the corresponding  $t=1$~Myr (it is the
youngest of the complexes discussed here).
The high expansion velocity estimated in this study is indicative of a shock effect. To verify this
hypothesis, we constructed the distribution of the  I(\SII 6717,6731)/I(\Ha) line ratio  based on the SCORPIO long-slit data (Sect.~\ref{sec_obs3}). Figure~\ref{fig:reg7-ls} shows the   distribution of this ratio along the slit  passing through the ring centre.

As is evident from Fig.~\ref{fig:reg7-ls}, the ratio \SIIHa\ $= 0.1-0.2$ in the central bright
nebula corresponds to  the recombination radiation of the \HII  region, but at the same time it increases
outwards and reaches   \SIIHa\ $= 0.35$ at the edges of the ring nebula, which suggests a possible shock wave effect.

The required kinetic energy input rate of $(2.5-9.9)\times10^{38}\ \mathrm{erg~s}^{-1}$ determined
above can be provided by winds from several WR or Of stars. Note that the ring nebula
is observed not inside the  \HI cloud, but against a background of tenuous diffuse gas. This suggests
that the ambient density of unperturbed gas calculated in  both methods is highly overestimated.
Given the uncertainty of the estimated $n_\mathrm{HI}$  in the neighbourhood of  Region \#~7,
even a single star with a strong wind would be enough. The search for WR stars conducted by
\citet{drissen93} produced no possible candidates in this region. We nevertheless consider it very
interesting to perform a detailed analysis of the stellar content of the  PI 01b cluster at the
centre of the ring nebula.

We cannot rule out the possibility that the large ring in Region \#~7 may be an old SNR.
At least,  with the ambient
density of $n_{\mathrm{HI}}\simeq 0.3~\mathrm{cm}^{-3}$ the ring nebula fits well the empirical evolutionary sequence of SNRs on the $\log(v_\mathrm{exp})$ versus $\log (R_{s}(n_{\mathrm{HI}})^{1/3})$ plot
suggested by \citet{lozinsk80a, lozinsk80b}. The expected radio surface brightness of an SNR 210~pc in diameter is about
$\Sigma_\mathrm{1GHz} = 10^{-22} - 10^{-21}~\mathrm{W m^{-2} Hz^{-1} sr^{-1}}$ (see the $\Sigma - D$ relation
in \citealt{lozinsk81, berezhko04, asvarov06}). This surface brightness
is less than the limiting brightness at  $\lambda=6$~cm in the observations by \citet{walter98}, which
were the basis for the identification of the SNR in Region \#~11.

Therefore,   the nature of the youngest star-forming complex in the SGS remains an open issue.

\begin{figure}
\includegraphics[width=0.752\linewidth, angle=90]{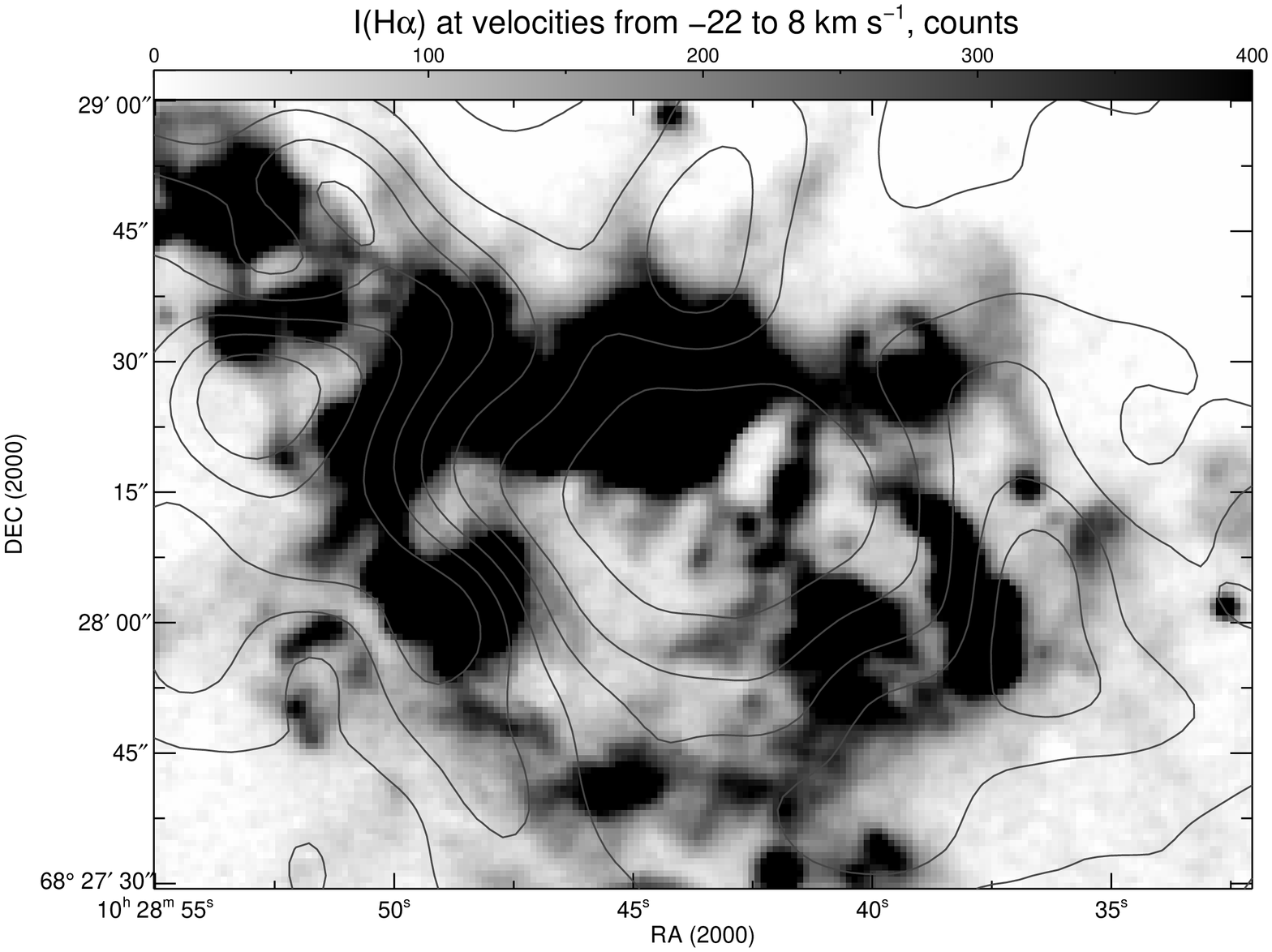}

\includegraphics[width=0.752\linewidth, angle=90]{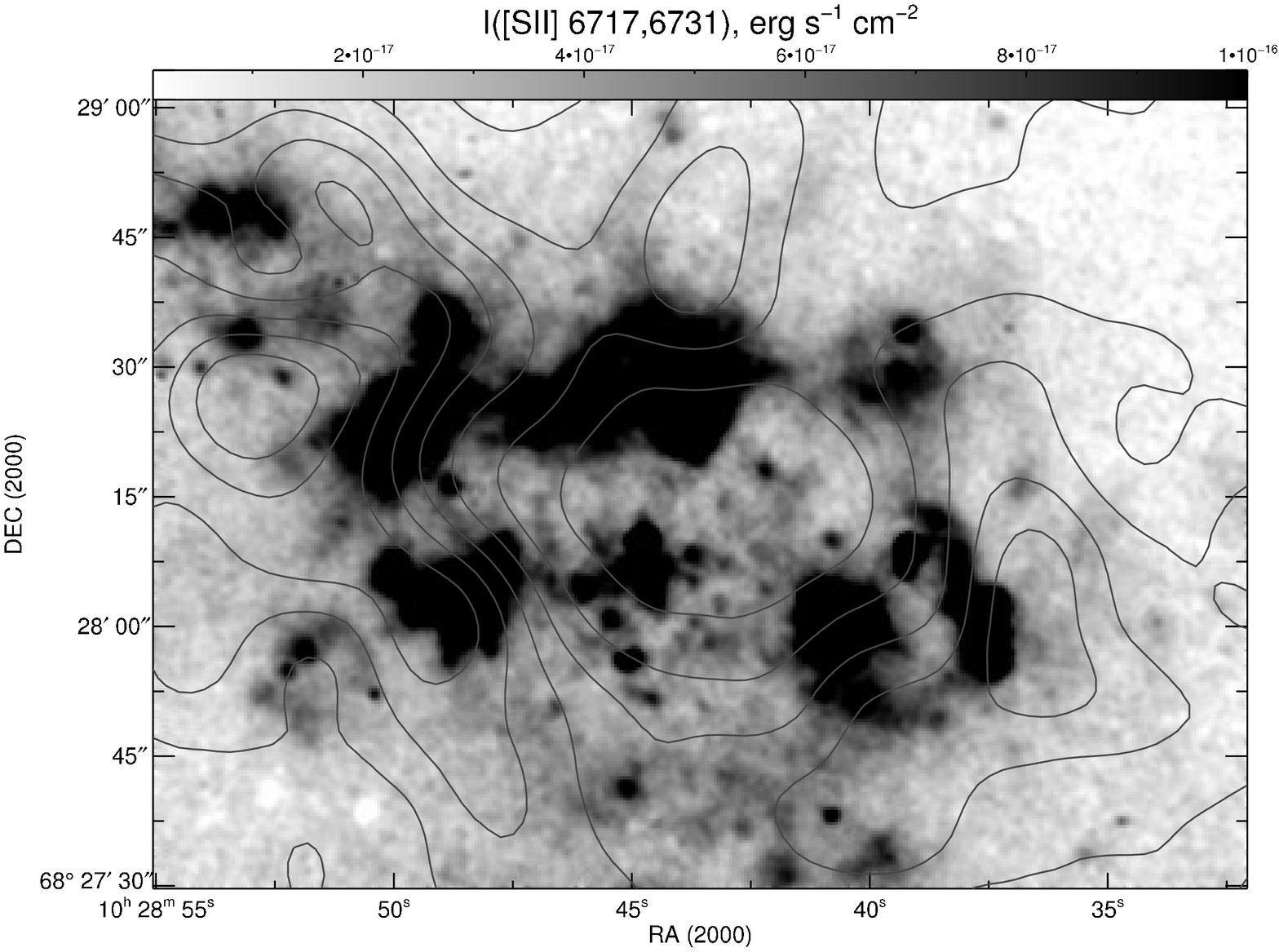}

\caption{\Ha map of the SGS region at velocities ranging from $-22 \kms$ to $8 \kms$ based on FPI
observations (top) and the narrow-band filter image in the \SII 6717, 6731 \AA\, lines (bottom)
with the $[0.5, 1.0, 1.5, 2.0, 2.5, 3.0, 3.5, 4.0]\times10^{21}$ cm$^{-2}$ \HI column density
contours overlaid.}
\label{fig:diffgas}
\end{figure}

\subsection{Diffuse ionized medium inside and around the SGS}\label{sec:HII-diffuse}

Deep ground-based images allowed us to identify a faint diffuse component of ionized gas in the
SGS region, which does not show in the \textit{HST}/ACS images.

Inside the SGS in the central Region \#~1 and also outside the SGS we observed \Ha line emission with an average  brightness level of $(6-8)\times10^{-17} \mathrm{erg~s}^{-1}\mathrm{cm}^{-2}$
and \SII line emission with an average brightness level
of $(2.5-3.5)\times10^{-17}\mathrm{erg~s}^{-1}\mathrm{cm}^{-2}$.

As is evident from Fig.~\ref{fig:siiha}, the \SIIHa\ intensity ratio is  enhanced both in the
border regions outside the  SGS and in the outer layers of Region \#~1 inside the SGS.

To analyse the kinematics of this faint diffuse emission, we constructed the \Ha line PV diagrams
for several regions outside the \HII complexes and several PV diagrams at the centre of
Region \#~1 (Fig.~\ref{fig:all-pv}, plots 15 -- 20).

As is evident from Fig.~\ref{fig:all-prof} (diff-2), the \Ha line of the faint diffuse emission in the outer SGS
regions has a single-component profile with an FWHM and peak velocity of
$40-45$ and  $-10 \kms$, respectively. These are typical parameters for the SGS region. However, there are several faint regions observed with kinematic evidence of local perturbations or a presence of an expanding shell (see, e.g., profile diff-1 in Fig.~\ref{fig:all-prof}).

The line profile of the central Region \#~1 shows a well-defined two-component structure corresponding to
the expansion of very faint diffuse  structures at a velocity of $v_\mathrm{exp} \simeq 33 \pm 2 \kms$.
The kinematic age of the central cluster in Region \#~1, which is responsible for the formation of
the  SGS, is equal to 14~Myr \citep{walter98, walter99}. Therefore the bright \HII nebula is
not observed at this location (\Ha arises from the recombination of gas ionized by the radiation of massive stars
 and is observed over their typical  lifetimes  $\le 10$ Myr).
 However, one can still see the effect of shocks  from
stellar winds and SN explosions originating from the central stellar complex in Region \#~1.  \citet{yukita12}
revealed a bright point-like source CXOU J102843.0+682816 in the SGS in the $0.3-8.0$~keV band and detected
 an excess of diffuse low surface brightness X-ray emission within and around the \HI hole
after removing the point source from the image (see their fig.~1).
According to \citet{sharma14}, superbubbles can retain a large fraction (up to $35-40 \%$
for $n_{\mathrm{HI}} = 1~\mathrm{cm}^{-3}$) of the initial energy obtained from SN explosions as bubble thermal energy
and kinetic energy of the shell. This reserve of mechanical energy powers the observed diffuse X-ray emission of the hot
gas that fills the central hole, drives the expansion of the neutral SGS at a velocity of  $25 \kms$,
and may also be responsible for the observed expansion of the faint ionized gas structures inside the
SGS at a velocity of   $\sim 33 \kms$.

The observed parameters of local diffuse gas in the SGS region are similar to those of the
extra-planar diffuse ionized medium (DIG) in spiral and Irr galaxies.
The observed  \SIIHa\ and \NIIHa\ line ratios in the DIG are higher than those in the classical \HII
regions. The ionization of DIG   is traditionally  explained by the leakage of ionizing photons
from the \HII regions \citep[see, e.g.,][and references therein]{seon09, hidgam06}.
To explain the increase of \NIIHa, \SIIHa\ and  \OIIIHb\ with galactic height, additional sources of
ionization were considered: ionization by shocks or turbulent mixing layers may be responsible for some of
the DIG emission.  Note that the photoionization simulations of a multi-component
interstellar medium by \citet{wood04} do not require as much additional heating  as the previous studies.

We can similarly explain the faint diffuse ionized gas emission observed in the SGS as a result of
leakage from the bright \HII regions of both the ionizing photons and mechanical energy of stellar winds
and supernovae. All the \HII complexes studied here indeed have a non-uniform, clumpy, or filamentary
structure,
which allows the radiation and shocks to leak outside through low-density regions.
In a number of cases the results of such a leakage are actually observed. One striking example are the `horns':
two faint emission structures observed beyond the northern boundary of Region \#~6 around a dense
\HI cloud (see Fig.~\ref{fig:diffgas}). These structures are observed both in the \Ha
and \SII lines.

\subsection{The subsequent evolution of the HI structure}

 The SGS in IC~2574 is one of the most dynamically active \HI structures among Irr galaxies, because we observe it during a period of violent star formation in its walls triggered by the expansion of the SGS. What future awaits this structure?

The most dynamically active shell-like \HII complexes exhibiting the expansion effect are regions \#~4, 6
and 7, which are located in the northern wall of the SGS. The expansion of Region \#~6 rapidly
disperses the
local \HI gas in the northern wall of the SGS. This is, in  particular, clearly demonstrated by the emission
features mentioned above  -- the `horns' located beyond the northern boundary of  Region \#~6
(see Fig~\ref{fig:diffgas}).

The destruction of the northern wall of the \HI supergiant shell due to \HII complexes will
result in the growth of the  SGS and, ultimately, in its merging with the neighbouring
H~{\sc i} structures. One must primarily expect the SGS to merge with the supergiant  \HI shells  \#~31
in the west and \#~37 in the north  \citep[see figs. 9 and 14 in][]{walter99}. As a result, after
several billion years there will be a system of giant adjoining and/or interacting shell-like
H~{\sc i} structures similar to those observed in the Irr galaxy Holmberg~II \citep[see][]{walter98}.


\section{Conclusions}\label{sec:conclusions}

\begin{enumerate}

\item To study the kinematics of ionized gas in the regions of triggered star formation in the walls of the SGS, we performed  observations  with the SAO RAS 6-m telescope  using SCORPIO and SCORPIO-2 multi-mode focal reducers operated in the scanning Fabry--Perot  interferometer, long-slit and direct imaging modes.

\item Using the THINGS survey data to analyse the kinematics of neutral gas, we `derotated' both (21 cm and H$\alpha$)  data cubes with a circular rotation model of IC~2574 constructed by fitting  the \HI  velocity field.

\item The consideration of global \HI kinematics confirms the SGS expansion velocity obtained earlier by \citet{walter99}. A more detailed analysis of the \HI kinematics revealed the non-uniform structure and kinematics of the SGS; it is shown to be located at the far side of the galactic disc plane.

\item We perform a detailed analysis  of the kinematics of all  extended  \HII complexes in the SGS. We estimated the expansion velocities for four regions, which allowed us to determine their kinematic ages and the kinetic energy input rate required to drive their formation. For the remaining  \Ha complexes we  estimated  the upper limit of the expansion velocity  and the corresponding limiting age and kinetic energy input. A comparison with the age and energy input from the stellar population of the complexes based on the data of \citet{stewart00} and \citet{yukita12} shows that the energy input is sufficient in all complexes except for Region \#~7.

\item We discuss in detail the nature of Region \#~7, which is a regular faint thin  ring with a bright inner region in its southern part. The obtained expansion velocity of $65~\kms$ shows that Region \#~7 is the youngest \Ha complex in the SGS ($t=1.0$~Myr).  The high expansion velocity is indicative of a shock. The significant role of shock waves in the gas ionization process is evident from the analysis of the  \SIIHa\, line  ratio.

\item  Region \#~11 was considered as a possible SNR based on radio observations \citep{walter98}.
We measured its expansion velosity and estimate the age of the SNR and the initial SN explosion energy $E_{o}$ under two assumptions: the adiabatic stage or intense radiative cooling.
In both cases the results confirm that Region \#~11 is an old remnant of a standard SN. Therefore the inferred
high expansion velocity and the high relative \SII line intensities provide further evidence that Region \#~11 is indeed an SNR.

\item  Our observations reveal  a faint diffuse component of ionized gas in the SGS area, which did not show up previously in \textit{HST}/ACS images. In the central Region \#~1, as well as outside the  SGS, we observe \Ha and \SII line emission with an average surface brightness of $(6-8)\times10^{-17} \mathrm{erg~s}^{-1}\mathrm{cm}^{-2}$ and $(2.5-3.5)\times10^{-17}\mathrm{erg~s}^{-1}\mathrm{cm}^{-2}$, respectively.
The \SIIHa\ intensity ratio is higher both in the border regions outside the  SGS and in the outer layers of Region \#~1. The observed parameters of local diffuse gas in the SGS region are similar to those of extra-planar diffuse ionized medium  in Sp and Irr galaxies. An analysis of the kinematics of this faint diffuse emission shows  a single-component line profile in the outer regions of the SGS, but reveals some structures that show perturbed kinematics with a multi-component \Ha line profile. The central Region \#~1 shows well-defined two-component line profiles corresponding to the expansion of faint diffuse inner structures with a velocity of  $\simeq 33 \kms$.
This expansion velocity and the soft X-ray diffuse emission within the \HI hole (and extending slightly beyond it) are extant traces of the initial effect of shock waves from stellar winds and SN explosions in the central stellar complex of  Region \#~1.

\item Based on the results of the SGS gas kinematics analysis we made speculative conclusions about the future of this structure: it will probably merge with the neighbouring supershells. The vicinity of the SGS will possibly look like the interstellar medium in Holmberg~II -- a typical example of an object with multiple interacting supergiant shells with triggered star formation at their rims.

\end{enumerate}

\section*{ACKNOWLEDGEMENTS}

We thank the anonymous referee for valuable comments which have improved the clarity of this paper.

This work was supported by the Russian Foundation for Basic Research (project nos 12-02-31356
and 14-02-00027) and partly supported by the  `Active Processes in Galactic and Extragalactic Objects' basic research program of the Department of Physical Sciences of RAS OFN-17. A.V. Moiseev acknowledges the support from the `Dynasty' Foundation.
We also thank Victor Afanasiev and Alexander Burenkov who performed the imaging and long-slit observations at the 6-m telescope, and Fabian Walter for providing us with the \HI data cube of IC~2574 and the corrected version of their age estimates of \HII complexes.

This work is based on observations obtained with the 6-m telescope of the Special Astrophysical Observatory of the Russian Academy of Sciences. The observations were carried out with the financial support of the Ministry of Education and Science of the Russian Federation (contract nos 16.518.11.7073 and 14.518.11.7070). We used observations made with the NASA/ESA \textit{Hubble Space Telescope} and obtained from the \textit{Hubble} Legacy Archive, which is a collaboration between the Space Telescope Science Institute (STScI/NASA), the Space Telescope European Coordinating Facility (ST-ECF/ESA) and the Canadian Astronomy Data Centre (CADC/NRC/CSA). This work has made use of THINGS, `The \HI Nearby Galaxy Survey'.

\label{lastpage}

\appendix

\section{Channel map from FPI H$\alpha$ data cube}

\begin{figure*}
\includegraphics[width=\linewidth]{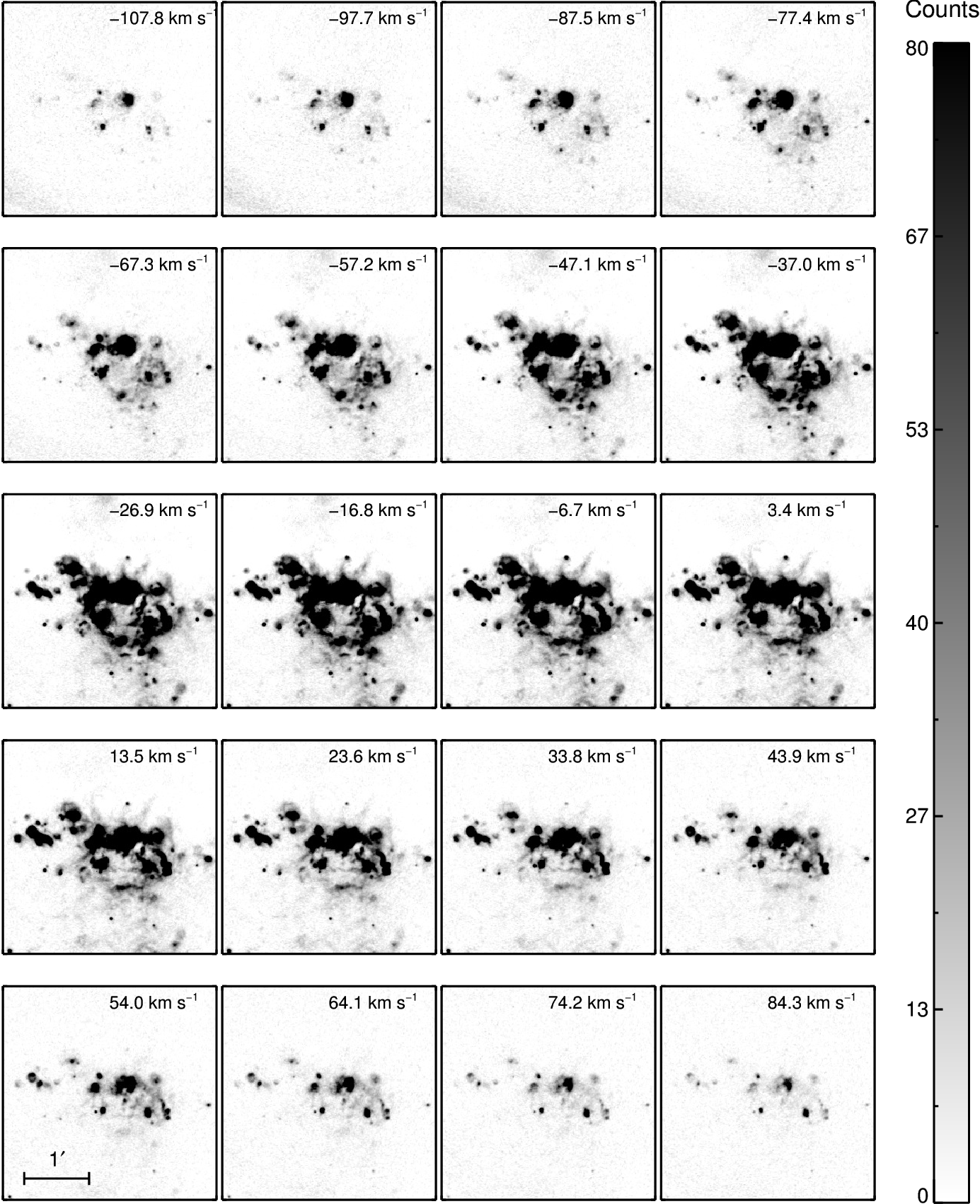}
\caption{Channel map from FPI H$\alpha$ data cube. The velocity of each channel is shown in the top right corner of the panel.}\label{fig:Ha-channel}
\end{figure*}

In this appendix we present a channel map of the `derotated' \Ha data cube of the SGS in IC~2574. During the observations we obtained a 40-channel data cube. We present in Fig.~\ref{fig:Ha-channel} the image of 32 channel map with noticeable \Ha intensity.

\end{document}